\documentclass[aps,pra,twocolumn,superscriptaddress]{revtex4-2}
\usepackage{amssymb}
\usepackage{amsmath}
\usepackage{bm}
\usepackage{graphicx}
\usepackage{subcaption}
\usepackage{float}
\usepackage[hyperfootnotes=false]{hyperref}
\usepackage{tabularx}
\usepackage{dcolumn}
\usepackage{isotope}
\usepackage{color}
\usepackage{siunitx}
\DeclareMathOperator{\sinc}{sinc}

\begin{document}

\title{Analyzing the collective emission of a Rydberg-blockaded single-photon source based on an ensemble of thermal atoms}

\author{Jan~A.~P.~Reuter}
\email[]{ja.reuter@fz-juelich.de}
\affiliation{Peter Grünberg Institute - Quantum Control (PGI-8), Forschungszentrum Jülich GmbH, Wilhelm-Johnen-Straße, 52428 Jülich, Germany}
\affiliation{Institute for Theoretical Physics, University of Cologne, Zülpicher Straße 77, 50937 Cologne, Germany}

\author{Max~Mäusezahl}
\email[]{mmaeusezahl@pi5.physik.uni-stuttgart.de}
\affiliation{5th Institute of Physics and Center for Integrated Quantum Science and Technology, University of Stuttgart, Pfaffenwaldring 57, 70569 Stuttgart, Germany}

\author{Felix~Moumtsilis}
\affiliation{5th Institute of Physics and Center for Integrated Quantum Science and Technology, University of Stuttgart, Pfaffenwaldring 57, 70569 Stuttgart, Germany}

\author{Tilman~Pfau}
\affiliation{5th Institute of Physics and Center for Integrated Quantum Science and Technology, University of Stuttgart, Pfaffenwaldring 57, 70569 Stuttgart, Germany}

\author{Tommaso~Calarco}
\affiliation{Peter Grünberg Institute - Quantum Control (PGI-8), Forschungszentrum Jülich GmbH, Wilhelm-Johnen-Straße, 52428 Jülich, Germany}
\affiliation{Institute for Theoretical Physics, University of Cologne, Zülpicher Straße 77, 50937 Cologne, Germany}
\affiliation{Dipartimento di Fisica e Astronomia, Università di Bologna, 40127 Bologna, Italy}

\author{Robert~Löw}
\affiliation{5th Institute of Physics and Center for Integrated Quantum Science and Technology, University of Stuttgart, Pfaffenwaldring 57, 70569 Stuttgart, Germany}

\author{Matthias~M.~Müller}
\affiliation{Peter Grünberg Institute - Quantum Control (PGI-8), Forschungszentrum Jülich GmbH, Wilhelm-Johnen-Straße, 52428 Jülich, Germany}

\date{\today}

\begin{abstract}
An ensemble of Rubidum atoms can be excited with lasers such that it evolves into an entangled state with just one collective excitation within the Rydberg blockade radius. The decay of this state leads to the emission of a single, antibunched photon. For a hot vapor of Rubidium atoms in a micro cell we numerically study the feasibility of such a single-photon source under different experimental conditions like the atomic density distribution and the choice of electronic states addressed by the lasers. For the excitation process with three rectangular lasers pulses, we simulate the coherent dynamics of the system in a truncated Hilbert space. We investigate the radiative behavior of the moving Rubidum atoms and optimize the laser pulse sequence accordingly. We find that the collective decay of the single-excitation leads to a fast and directed photon emission and further, that a pulse sequence similar to a spin echo increases the directionality of the photon. Finally, we analyze the residual double-excitations and find that they do not exhibit these collective decay properties and play only a minor deleterious role.
\end{abstract}

\maketitle

\section{Introduction}\label{Introduction}
Single-photon sources are essential components for many upcoming technologies including quantum computation \cite{Brien2007,Kok2007} and quantum communication \cite{Gisin2007,Cavaliere2020}. Most existing single-photon source platforms use single emitters that naturally emit antibunched light, e.g. NV-centers \cite{Mizuochi2012}, quantum dots \cite{Senellart2017}, single atoms \cite{Hijlkema2007}, or single ions \cite{Barros2009}, all of them with specific advantages and disadvantages \cite{Eisaman2011,Lounis2005}. Here, instead, we consider ensembles of Rubidum atoms and leverage the Rydberg blockade effect \cite{Heidemann2007} to generate an entangled many-body state with only a single excitation based on the van-der-Waals interaction \cite{Baluktsian2013} between the atoms. The strong Rydberg-Rydberg interaction makes Rydberg atoms interesting for several applications in the field of quantum technologies \cite{Moelmer2010}, e.g., for entanglement generation or phase gates \cite{Jaksch2000,Gaetan2009,Isenhower2010,Wilk2010,Mueller2014}, quantum simulations~\cite{Labuhn2016,Barredo2018,Scholl2021} or single-photon sources~\cite{Saffman2002,Pedersen2009,Mueller2013,Dudin2012,Bariani2012,Ripka2018}. While such a single-photon source has been first realized with ultra-cold atoms~\cite{Dudin2012}, it has been shown, that cooling might be omitted and a single-photon source with Rydberg atoms can also operate at room temperature~\cite{Mueller2013,Ripka2018}. This, however, implies driving the atoms off-resonantly due to their individual Doppler shift, a finite time of flight, and movement-induced decoherence. Part of these challenges could be mitigated by an optimal pulse sequence to render the excitation process robust against this motional noise~\cite{Mueller2013,Mueller2022}.

In this paper we build on the work of Refs.~\cite{Mueller2013,Ripka2018,Reuter2022} and further investigate the emission behavior of single-photon sources based on room-temperature Rubidum-atom ensembles. We also investigate the influence of the excitation pulses on the phase information encoded in the entangled excited state which influences the directionality of the photon emission. Furthermore, we will also discuss the behavior of double-excitations and the emission of two photons. This has already been done by \cite{Bariani2012}, albeit with a different approach and in a slightly different context. We start by describing the setup and our model of the system in section~\ref{System Setup and Model}. In section~\ref{Numerical methods} we discuss our numerical approach to simulate the quantum many-body dynamics of the atomic ensemble. Finally, in section~\ref{Results} we present and discuss the results of the simulations both for the emission of a single-photon as well as a two-photon emission.

\section{System Setup and Model}\label{System Setup and Model}
\begin{figure}[b]
\begin{subfigure}{0.75\linewidth}
	\includegraphics[width=\linewidth]{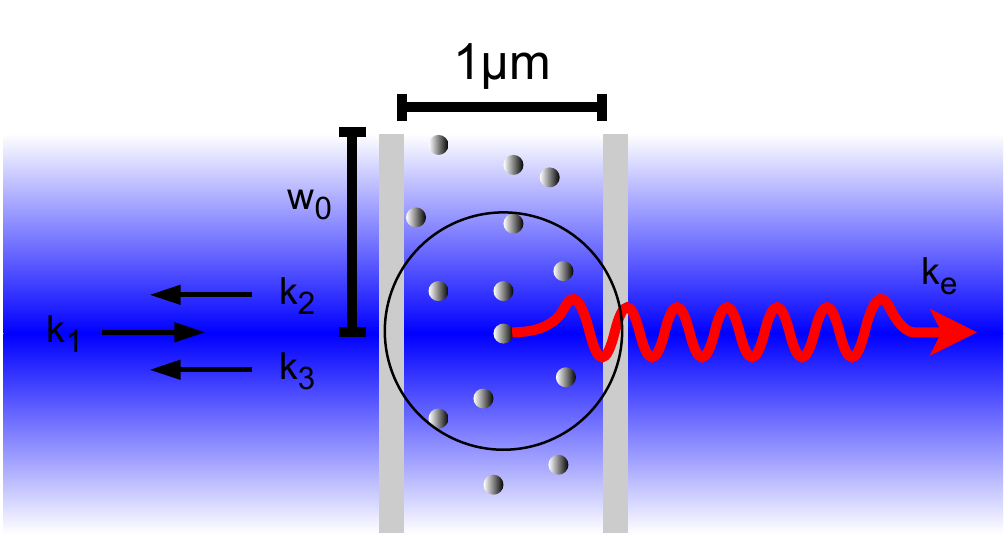}
	\caption{} \label{Setup}
\end{subfigure}
\begin{subfigure}{\linewidth}
	\includegraphics[width=0.5\linewidth]{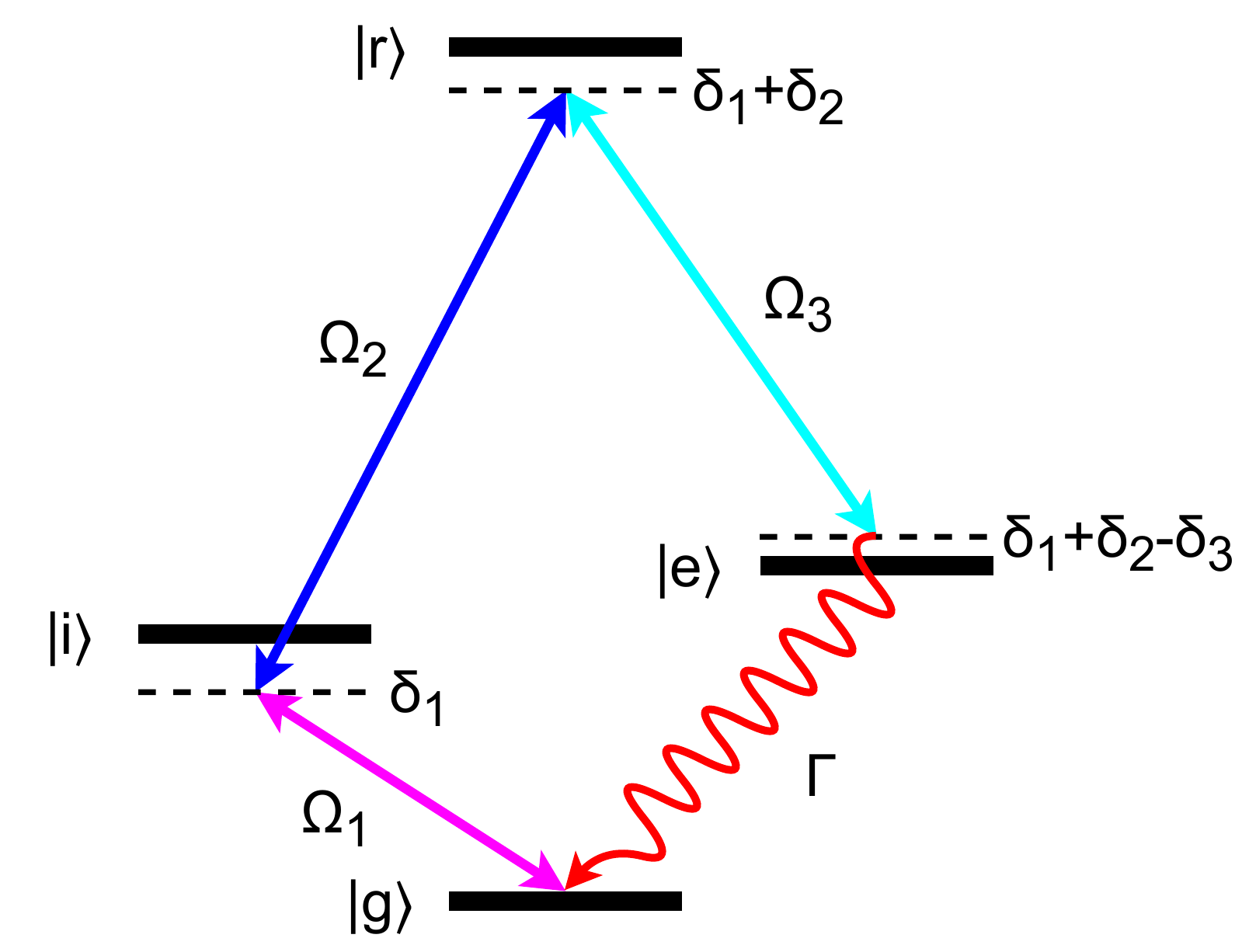}
	\caption{} \label{Level_scheme}
\end{subfigure}
\caption{Setup of the system. In (a) one can see \isotope[85]{Rb} atoms in a vapor cell which get excited by three anti-parallel lasers. The cell length as well as the laser beam waist $\mathrm{w}_0$ has the size of the Rydberg blockade radius. Instead, (b) shows the 4-level scheme with the respective Rabi frequencies $\Omega_j$ and atom decay rate $\Gamma$.}
\label{Setup and Level_scheme}
\end{figure}

We consider $N$ neutral \isotope[85]{Rb} atoms in a microscopic vapor cell at room temperature as shown in Fig.~\ref{Setup}. In one direction the atoms are confined by the cell walls that have a distance of \SI{1}{\micro\meter}. Three lasers couple the ground state $\vert g \rangle=\vert 5 S_{1/2} \rangle$ to the Rydberg state $\vert r \rangle=\vert 40 S_{1/2} \rangle$ via an intermediate state $\vert i \rangle=\vert 5 P_{1/2} \rangle$, and finally to the excited state $\vert e \rangle=\vert 5 P_{3/2} \rangle$ (Fig.~\ref{Level_scheme}). This procedure is also known as four-wave mixing \cite{Koelle2012,Ripka2016} where $\mathbf{k}_0=\mathbf{k}_1+\mathbf{k}_2-\mathbf{k}_3$ is the mixed wave vector of the three incoming lasers.
For the transition from the ground to the Rydberg state we choose the first laser to be red-detuned by \SI{100}{\giga\hertz} and the second laser blue-detuned also by \SI{100}{\giga\hertz}, so that the two-photon transition is resonant again. This procedure minimizes the amount of population in the intermediate state and makes the two-photon transition more robust.
The resonant transition frequencies have been taken from experimental data \cite{Steck2021,Glaser2020} for the $\vert g \rangle \leftrightarrow \vert i \rangle,\vert e \rangle$ transitions and calculated using the ARC-Alkali-Rydberg-Calculator \cite{Arc} for $\vert i \rangle,\vert e \rangle \leftrightarrow \vert r \rangle$.

The atoms are initially prepared in the ground state $\vert G \rangle:=\vert g_1... g_N\rangle$ with the goal to exploit the Rydberg blockade to create an entangled state which contains exactly one single excitation and stores the directional information of the incoming photons in the relative phases. This entangled state is the W-state $\vert W (t_W) \rangle= \sum_n w_n e^{i\mathbf{k}_0 \cdot \mathbf{R}_n(t_W)} \vert e_n \rangle$, where $\mathbf{R}_n(t)$ is the position of atom $n$ at time $t$, $t_W$ depends on the laser pulses as will become clear in section~\ref{Results}, and the state $\vert e_n \rangle:=\vert g_1... e_n ... g_N\rangle$ denotes a single excitation of atom $n$ in state $|e\rangle$ with all other atoms in the ground state $|g\rangle$, and the coefficients $w_n$ depend on the distribution of the atoms and on the laser pulses.
When this W-state, which is similar to a discrete Fourier transformation of the mixed photon wave $\mathbf{k}_0$, decays, a single-photon with wave vector $\mathbf{k} \approx \frac{k_e}{k_0}\mathbf{k}_0$ is emitted in the preferred direction $\mathbf{k}_0$ and with a frequency close to the resonat transition frequency to the ground state $\omega_e=c k_e$.
For the laser geometry, we decide to set the second and third laser anti-parallel to the first one. This minimizes the detuning of the lasers from the Doppler shift and the dephasing during the decay process \cite{Mueller2013,Reuter2022}.

The working principle of the single-photon source can be divided into two temporally subsequent steps: the excitation process from the atomic ground state $\vert g \rangle$ to the excited state $\vert e \rangle$ via photon absorption from the three lasers and the decay process back to the ground state via photon emission. This separation is valid as long as the life time $\tau$ of the excited state is much longer than the pulse duration $t_0$. This is given in our case, since the excited state has a life time of $\tau=\SI{26.2}{\nano\second}$ \cite{Steck2021}, while we consider pulse durations around $t_0\approx \SI{1.5}{\nano\second}$. We will thus model the two processes separately and connect them by using the final state of the excitation process as initial state of the decay process.

\subsection{The excitation process}\label{The excitation process}
The many-body Hamiltonian in the lab frame $\hat{H}(t)$ contains the eigenenergies of the single atom Hamiltonian $\hat{H}_{0}=\sum_s E_s \vert s \rangle \langle s \vert$, $s\in\{g,i,r,e\}$, the interaction $\hat{H}_{\mathrm{trans}}(t)=\sum_{n,j}\mathrm{e}\hat{\mathbf{\mathrm{r}}}_n \mathbf{\mathcal{E}}_{j,n}(t)$ between the single valence electron with dipole moment $\mathrm{e}\hat{\mathbf{\mathrm{r}}}_n$ of atom $n$ with a laser field $j$, and the atom-atom interaction $\hat{H}_{\mathrm{int}}(t)$.
For the excitation process we describe the system in a semi-classical way, which means that we define the electric field as a classical oscillating wave $\mathbf{\mathcal{E}}_{j,n}(t)=\mathcal{E}_j(t) \cos (\mathbf{k}_j \cdot \mathbf{R}_n(t)-\omega_j t) \mathbf{\epsilon}_j$ with linear polarization vector $\mathbf{\epsilon}_j$, while we treat the electronic states as a quantum system. To simplify the calculation, we transfer the system from the lab frame to the rotating frame by a unitary transformation $\vert \psi(t) \rangle=\hat{T}(t) \vert \tilde{\psi}(t) \rangle$. We choose $\hat{T}(t)=\exp (-\frac{it}{\hbar}(\hat{H}_{0}-\hat{\tilde{H}}_{0}))$ such that the drift part of the transformed Hamiltonian
$\hat{\tilde{H}}(t)=\hat{\tilde{H}}_{0} + \hat{\tilde{H}}_{\mathrm{trans}}(t) + \hat{H}_{\mathrm{int}}(t)$ takes on the form
\begin{eqnarray}
\hat{\tilde{H}}_{0} &&= -\hbar \sum_n \delta_{1,n} \vert i_n \rangle \langle i_n \vert 
+(\delta_{1,n} + \delta_{2,n}) \vert r_n \rangle \langle r_n \vert \nonumber \\
&&+(\delta_{1,n} + \delta_{2,n} - \delta_{3,n})\vert e_n \rangle \langle e_n \vert .
\label{eq1}
\end{eqnarray}
We then apply the Rotating Wave Approximation (RWA), which neglects the fast oscillating terms from the electric field on time scales of femtoseconds and only keeps oscillations in the range of pico- up to nanoseconds. It is important to notice that one needs to transform the system back to the lab frame after the simulation to get the actual particle-dependent phases of the quantum state.
The drift Hamiltonian in the rotating frame $\hat{\tilde{H}}_{0}$ contains the detunings $\delta_{j,n}=\delta_{j}-\mathbf{k}_j \cdot \mathbf{v}_n$ of the different laser frequencies with respect to the resonant transitions frequencies (e.g. $\delta_1=\omega_1-\frac{E_i-E_g}{\hbar}$), where we also take the Doppler shift of each particle into account. The single-atom off-diagonal elements
\begin{eqnarray}
\hat{\tilde{H}}_{\mathrm{trans}}(t)= \frac{\hbar}{2} \sum_{n}&& 
\Omega_{1,n}(t)   \vert i_n \rangle \langle g_n \vert 
+\Omega_{2,n}(t) \vert r_n \rangle \langle i_n \vert \nonumber\\
&&+\Omega_{3,n}(t) \vert r_n \rangle \langle e_n \vert 
+h.c.
\label{eq2}
\end{eqnarray}
given by the individual Rabi frequencies $\Omega_{j,n}(t)=\Omega_{j}(t) \exp \left( i \mathbf{k}_j \cdot \mathbf{R}_n(0) -\frac{\vert \mathbf{R}_{n}^\perp(t)\vert^2}{\mathrm{w}_{0,j}^2} \right)$, are determined by the laser intensity (e.g. $\Omega_1(t)=\frac{\mathrm{e}}{\hbar} \mathcal{E}_1(t) \langle i \vert \hat{\mathbf{\mathrm{r}}} \vert g \rangle \cdot \mathbf{\epsilon}_1$), the starting position of each atom, the time dependent position in the plane orthogonal to the laser direction $\mathbf{R}_{n}^\perp$, as well as the beam waist $\mathrm{w}_0$ of the Gaussian laser profile, orthogonal to its direction of propagation. Finally, the interaction energy, which doesn't change under the transformation, between two Rydberg atoms $n$, $m$ is given by
\begin{eqnarray}
\hat{H}_{\mathrm{int}}(t)= \sum_{n < m} \frac{C_6}{d_{n,m}^6(t)} \vert r_n r_m \rangle \langle r_n r_m \vert
\label{eq3}
\end{eqnarray}
including the distance between the two atoms $d_{n,m}(t)=\vert \mathbf{R}_n(t)-\mathbf{R}_m(t)\vert$ and coefficients $C_6=h \cdot \SI{642.1}{\mega\hertz\per\micro\meter\tothe{6}}$ \cite{Ripka2019}.

\subsection{The decay process}\label{The decay process}
We start again with the Hamiltonian in the lab frame $\hat{H}_\mathrm{d}(t)=\hat{H}_{\mathrm{d},0} + \hat{H}_{\mathrm{d,trans}}(t)$, where $\hat{H}_{\mathrm{d},0}$ contains the transition energies and $\hat{H}_{\mathrm{d,trans}}(t)$ is the interaction between the atoms and the electromagnetic field. Since we focus on the decay of the excited state we treat the atoms as two-level systems ($|g\rangle$ and $|e\rangle)$ and neglect any atom-atom interaction. To describe the decay process of the excited atoms and the photon emission we have to quantize the electromagnetic field $\mathbf{\mathcal{E}}_{n}(t)=\sum_{\mathbf{k},\mu} \mathcal{E}_k \left(a_{\mathbf{k},\mu}e^{i \mathbf{k}_j \cdot \mathbf{R}_n(t)}+a^\dagger_{\mathbf{k},\mu} e^{-i \mathbf{k}_j \cdot \mathbf{R}_n(t)}\right) \mathbf{\epsilon}_{\mathbf{k},\mu}$ with polarization $\mathbf{\epsilon}_{\mathbf{k},\mu}$ ($\mu \in \{1,2\}$) and annihilation operator $a_{\mathbf{k},\mu}$. Consequently, the transition elements are $\hat{H}_{\mathrm{d},0}=\hbar \omega_e \sum_n \vert e_n \rangle \langle e_n \vert+ \sum_{\mathbf{k},\mu} \hbar \omega_{\mathbf{k},\mu} a_{\mathbf{k},\mu}a^\dagger_{\mathbf{k},\mu}$. After transforming the system into the rotating frame via $\hat{T_\mathrm{d}}(t)=\exp (-\frac{it}{\hbar}\hat{H}_{\mathrm{d},0})$ and applying the RWA the Hamiltonian is
\begin{eqnarray}
\hat{\tilde{H}}_\mathrm{d}(t) = \hbar \sum_{n,\mathbf{k},\mu} \mathrm{g}_{\mathbf{k},\mu} \sigma_n^\dagger  a_{\mathbf{k},\mu}  e^{i\bm(\mathbf{k} \cdot \mathbf{R}_n(t) - (\omega_k-\omega_e) t\bm)} + h.c.
\label{eq4}
\end{eqnarray}
with the coupling constant $\mathrm{g}_{\mathbf{k},\mu}=\frac{\mathrm{e}}{\hbar}\mathcal{E}_k \langle e \vert \hat{\mathbf{\mathrm{r}}} \vert g \rangle \cdot \mathbf{\epsilon}_{\mathbf{k},\mu}$ between the atoms and the electromagnetic field and the lowering operator $\sigma_n = \vert g_n \rangle \langle e_n \vert$.

\subsubsection{Single-excitation and single-photon emission} \label{Single-exitation and single-photon emission}
We first focus on the collective decay of the singly-excited states $\vert e_n \rangle$ to the many-body ground state  $\vert G \rangle$ and the emission of a single-photon $\vert 1_{\mathbf{k},\mu} \rangle$. We thus consider a general state
$\vert \psi_e(t) \rangle = \sum_n \alpha_n(t)\vert e_n \rangle \vert 0 \rangle + \sum_{\mathbf{k},\mu} \beta_{\mathbf{k},\mu}(t) \vert G \rangle \vert 1_{\mathbf{k},\mu} \rangle$ 
with one shared excitation between the atoms and the single-photon modes, where the $\alpha_n(t)$ are the coefficients for the atomic excitation and the $\beta_{\mathbf{k},\mu}(t)$ for the photon modes, respectively. At $t=t_0$, right after the laser pulses, we consider the photon mode not to be occupied, i.e., $\beta_{\mathbf{k},\mu}(t_0)=0$. The Schrödinger equation then leads to two coupled differential equations
\begin{eqnarray}
i \dot{\alpha}_n(t)=\sum_{\mathbf{k},\mu} \mathrm{g}_{\mathbf{k},\mu} \beta_{\mathbf{k},\mu}(t) e^{i\bm(\mathbf{k} \cdot \mathbf{R}_n(t) - (\omega_k-\omega_e) t\bm)},
\label{eq5}
\end{eqnarray}

\begin{eqnarray}
i \dot{\beta}_{\mathbf{k},\mu}(t)=\sum_m \mathrm{g}_{\mathbf{k},\mu}^* \alpha_m(t) e^{-i\bm(\mathbf{k} \cdot \mathbf{R}_m(t) - (\omega_k-\omega_e) t\bm)}
\label{eq6}.
\end{eqnarray}

We follow the calculations of \cite{Pedersen2009,Mueller2013} by integrating Eq.~(\ref{eq6}) in time and insert it into (\ref{eq5}). We then apply the Wigner-Weisskopf approximation \cite{Scully1997,Jacobs2006} which assumes only small deviations of $\omega_k$  to be present around $\omega_e$ (i.e., the photon line width is narrow compared to its central frequency). Then, using the relation
$\int_0^\infty e^{-i(\omega_k-\omega_e)(t-t')} dk \approx \frac{2\pi}{c} \delta(t-t')$ leads to a finite number of coupled differential equations for the coefficients of the atomic excitation with the single-atom decay rate  $\Gamma=\frac{1}{\tau}$
\begin{eqnarray}
\dot{\alpha}_n(t)\approx - \frac{\Gamma}{2} \sum_m  \alpha_m(t) \sinc \bm(k_e d_{n,m}(t)\bm).
\label{eq7}
\end{eqnarray}

We are now interested in the population of the photon modes which have a momentum $\mathbf{k}$ similar to $\mathbf{k}_0$. The standard approach is to insert the solutions of $\alpha_n(t)$ into  Eq.~(\ref{eq6}) and sum over a discrete number of $\beta_{\mathbf{k},\mu}(t)$ in the 3D momentum space up to a certain angle \cite{Mueller2013,Pedersen2009}. Another approach which we want to introduce, is multiplying the time-integrated Eq.~(\ref{eq6}) with its complex conjugate. We then integrate over the whole momentum space and use the Wigner-Weisskopf approximation again to eliminate one time integral. Furthermore, we define the angle $\theta$ between $\mathbf{k}$ and $\mathbf{k}_0$ and $\phi$ between $\mathbf{k}^\perp$ and $\mathbf{d}_{n,m}^\perp(t)$, both being the orthogonal part of the previously defined vectors with respect to $\mathbf{k}_0$. For a detailed calculation see Appendix~\ref{Appendix A}. This results in $\sum_{\mathbf{k},\mu} \vert \beta_{\mathbf{k},\mu}(t) \vert^2 \approx \int_0^\pi p(\theta,t) d \theta$ with the population density function $p(\theta,t)$ given by  
\begin{widetext}
\begin{eqnarray}
p(\theta,t) =\frac{\Gamma}{2} \sin(\theta) \int\limits_{t_0}^t \sum_{n,m} \alpha_n(t') \alpha_m^*(t')  e^{-ik_e d_{n,m}^{\; \parallel}(t') \cos(\theta)}
J_0\bm(k_e d_{n,m}^\perp(t') \sin(\theta)\bm) \, dt',
\label{eq8}
\end{eqnarray}
\end{widetext}
where $J_0(\varphi)=\frac{1}{2\pi}\int_0^{2\pi}e^{-i\varphi \cos (\phi)} d\phi$ is the Bessel function of first kind. We can then split this function into two parts: the geometrical factor $\sin (\theta)$, which is the radius of an infinitesimal slice of a unit sphere around the ensemble, and the sum over coupled singly-excited coefficients, which contains the information on the atoms and on the absorbed photon. The advantage is that we now only have to discretize in the angle $\theta$ to calculate the photon emission numerically. Furthermore, the total probability of emitted photon and remaining excitation is constant $\int_0^\pi p(\theta,t) d \theta+\sum_n \vert \alpha_n(t) \vert^2=\sum_n \vert \alpha_n(t_0) \vert^2$ and
as a consequence, the collective decay rate can be conveniently calculated from the coefficients for the atomic excitation: $\frac{\partial}{\partial t}\int_0^\pi p(\theta,t) d \theta = - \frac{\partial}{\partial t}\sum_n \vert \alpha_n(t) \vert^2$.
\subsubsection{Double-excitation and two-photon emission}\label{Double-excitation and two-photon emission}
In more detail, after the excitation process, the population is distributed not only over a singly-excited state and some residual ground state population as considered in the previous subsection. The main additional contribution is the undesired population of doubly-excited states $\vert e_n, e_m \rangle$ (i.e., all atoms in the ground state except for atoms $m$ and $n$), which can lead to the emission of two photons and thus are potentially detrimental for the utilization of a single-photon source. Therefore, we want to investigate the decay of such doubly-excited states and the emission of the two photons. We consider the state $\vert \psi_{ee}(t) \rangle = \sum_{n < m} \alpha_{n,m}(t)\vert e_n, e_m \rangle \vert 0 \rangle +  \sum_{n,\mathbf{k},\mu} \beta_{n,\mathbf{k},\mu}(t) \vert e_n \rangle \vert 1_{\mathbf{k}, \mu} \rangle +\sum_{\mathbf{k},\mu, \mathbf{q},\eta} \gamma_{\mathbf{k},\mu,\mathbf{q},\eta}(t) \vert G \rangle \vert 1_{\mathbf{k}, \mu}, 1_{\mathbf{q}, \eta} \rangle$, where the $\alpha_{n,m}(t)$ are the coefficients for the double-excitations of the atom, $\beta_{n,\mathbf{k},\mu}(t)$ for the states with one atomic and one photonic excitation, and $\gamma_{\mathbf{k},\mu,\mathbf{q},\eta}(t)$ for the combinations of two photons with all the atoms in the ground state. Here, we neglect the fact that photons are indistinguishable, but distinguish between first $\vert 1_{\mathbf{k}, \mu} \rangle$ (with wave vector $\mathbf{k}$ and polarization $\mu$) and second $\vert 1_{\mathbf{q}, \eta} \rangle$ photon (with wave vector $\mathbf{q}$ and polarization $\eta$), following Ref.~\cite{Scully1997}. The Schrödinger equation then leads to three coupled differential equations
\begin{eqnarray}\label{eq10}
i\dot{\alpha}_{n,m}(t)&&=\sum_{\mathbf{k},\mu} \mathrm{g}_{\mathbf{k},\mu} \beta_{n,\mathbf{k},\mu}(t)e^{i\bm(\mathbf{k} \cdot \mathbf{R}_m(t) - (\omega_k-\omega_e) t\bm)}\\
&&+\sum_{\mathbf{k},\mu} \mathrm{g}_{\mathbf{k},\mu} \beta_{m,\mathbf{k},\mu}(t) e^{i\bm(\mathbf{k} \cdot \mathbf{R}_n(t) - (\omega_k-\omega_e) t\bm)}\,, \nonumber
\end{eqnarray}
\begin{eqnarray}\label{eq11}
i\dot{\beta}_{n,\mathbf{k},\mu}(t)&&=\sum_{l \neq n} \mathrm{g}_{\mathbf{k},\mu}^* \alpha_{n,l}(t) e^{-i\bm(\mathbf{k} \cdot \mathbf{R}_l(t) - (\omega_k-\omega_e) t\bm)}\\
&&+ \sum_{\mathbf{q},\eta} \mathrm{g}_{\mathbf{q},\eta} \gamma_{\mathbf{k},\mu,\mathbf{q},\eta}(t) e^{i\bm(\mathbf{q} \cdot \mathbf{R}_n(t) - (\omega_q-\omega_e) t\bm)}\,,\nonumber
\end{eqnarray}
\begin{eqnarray}\label{eq12}
i\dot{\gamma}_{\mathbf{k},\mu,\mathbf{q},\eta}(t)&&=\sum_l \mathrm{g}_{\mathbf{q},\eta}^* \beta_{l,\mathbf{k},\mu}(t)e^{-i\bm(\mathbf{q} \cdot\mathbf{R}_l(t) 
- (\omega_q-\omega_e) t\bm)} \,.
\end{eqnarray}
Corresponding to the product rule, we then make the ansatz $\beta_{n,\mathbf{k},\mu}(t)=\beta_{n,\mathbf{k},\mu}^\alpha(t)\cdot\beta_{n,\mathbf{k}}^\gamma(t)$ and associate with the derivative $\dot{\beta}_{n,\mathbf{k},\mu}^\alpha(t)\cdot\beta_{n,\mathbf{k}}^\gamma(t)$ the incoming population from the double-excitations given by the first term in Eq.~(\ref{eq11}) and with 
$\beta_{n,\mathbf{k},\mu}^\alpha(t) \cdot \dot{\beta}_{n,\mathbf{k}}^\gamma(t)$ the outgoing population into the two-photon state given by the second term. This is close to the calcuation in \cite{Scully1997}, but generalized to a many-body state. We can then decouple the double-excitations in a similar way as in the single-excitation case in Eq.~(\ref{eq7}) by inserting $\beta_{n,\mathbf{k},\mu}(t)=-i\sum_{l \neq n} \mathrm{g}_{\mathbf{k},\mu}^* \int \limits_{t_0}^t \alpha_{n,l}(t')\frac{\beta_{n,\mathbf{k}}^\gamma(t)}{\beta_{n,\mathbf{k}}^\gamma(t')} e^{-i\bm(\mathbf{k} \cdot \mathbf{R}_l(t') - (\omega_k-\omega_e) t'\bm)} dt'$ into Eq.~(\ref{eq10}) and obtain
\begin{eqnarray}
\dot{\alpha}_{n,m}(t)\approx &&-\frac{\Gamma}{2} \sum_{l \neq n} \alpha_{n,l}(t) \sinc \bm(k_e d_{m,l}(t)\bm) \nonumber \\
&&-\frac{\Gamma}{2} \sum_{l \neq m} \alpha_{m,l}(t) \sinc \bm(k_e d_{n,l}(t)\bm).
\label{eq13}
\end{eqnarray}

Furthermore, we can also decouple the remaining excitation and first-photon states from the two-photon states by inserting the time integral of Eq.~(\ref{eq12}) into Eq.~(\ref{eq11}) and following the same calculation and get:
\begin{eqnarray}
\dot{\beta}_{n,\mathbf{k},\mu}(t) \approx &&-i\sum_{l \neq n} \mathrm{g}_{\mathbf{k},\mu}^* \alpha_{n,l}(t) e^{-i\bm(\mathbf{k} \cdot \mathbf{R}_l(t) - (\omega_k-\omega_e) t\bm)} \nonumber \\
&&-\frac{\Gamma}{2}\sum_l \beta_{l,\mathbf{k},\mu}(t) \sinc \bm(k_e d_{n,l}(t)\bm).
\label{eq14}
\end{eqnarray}

Also here, we can calculate a photon-population density function for the first photon (see also Appendix~\ref{Appendix A}) and in principle, also for the second photon. Nevertheless, we will not present the formula for the second photon since, unfortunately, it is in any case not feasible to calculate it numerically (as we will do in Sec.~\ref{Results} for Eq.~(\ref{eq8}) and (\ref{eq9})) since we would have to sum over all particle pairs twice leading to an order of $\mathcal{O}(N^4)$ terms and double time-integrals over them. The density function for the first photon is given by
\begin{widetext}
\begin{eqnarray}
p_2(\theta,t) =\frac{\Gamma}{2} \sin(\theta) \int\limits_{t_0}^t \sum_n \sum_{m,l \neq n} \alpha_{n,m}(t') \alpha_{n,l}^*(t')  e^{-ik_e d_{m,l}^{\; \parallel}(t') \cos(\theta)}
J_0\bm(k_e d_{m,l}^\perp(t') \sin(\theta)\bm) \, dt'.
\label{eq9}
\end{eqnarray}
\end{widetext}

However, we can approximate the emission rate of the second photon by integrating over all first photon modes, where $\beta_{n,k_e}^{\gamma}(t)$ is the mean value of $\beta_{n,\mathbf{k}}^{\gamma}(t')$ over all directions:
\begin{eqnarray}
&&\sum_{\mathbf{k},\mu,\mathbf{q},\eta} \partial_t \vert \gamma_{\mathbf{k},\mu,\mathbf{q},\eta} (t) \vert^2= \Gamma \sum_{n} \sum_{\substack{m \neq n \\ l \neq n}} \\ 
&& \int \limits_{t_0}^t \alpha_{n,m}(t')\alpha_{n,l}^*(t') \frac{\partial_t \vert \beta_{n,k_e}^\gamma(t) \vert^2 }{\vert \beta_{n,k_e}^{\gamma}(t') \vert^2} \sinc \bm(k_e d_{m,l}(t')\bm) dt'. \nonumber
\label{eq15}
\end{eqnarray}

\section{Numerical methods}\label{Numerical methods}
If we want to numerically solve the equations that we have derived in the previous section, the main challenge is the quantum many-body dynamics of the excitation process. We have to efficiently sample the atoms and find a way to effectively truncate the Hilbert space for the excitation process. The solution of the decay process instead can be found with a standard differential equation solver.

\subsection{Particle sampling and velocity distribution}\label{Particle sampling and velocity distribution}
Before we simulate the excitation of the atoms via the three laser pulses, we have to sample the atoms with one of two possible distributions.
The first is the standard Maxwell-Boltzmann distribution at $T=200^\circ$C, where we assume random starting positions and a Gaussian velocity distribution in each direction. The second is produced by light-induced atomic desorption (LIAD) \cite{Meucci1994,Christaller2022} where a completely off-resonant laser pulse releases atoms which are sticking at the cell walls. For the distribution in the LIAD case, these atoms are emitted orthogonal to the glass cell walls (here also parallel to the desorption laser) leading to a directional velocity distribution $P(v,\theta)=av^2 \exp (-\frac{v^2}{b^2}) \cos (\theta)$, where we choose the parameters $a=\SI{1.1e-7}{\second^3\per\meter^3}$ and $b=\SI{271}{\meter\per\second}$ according to reference~\cite{Christaller2022}. With that, one can calculate the time-dependent average number of atoms which have not collided into the cell walls $\langle N \rangle (t)=N_0[1-\exp (-(\frac{\Delta x}{bt})^2)]$ if we expect that all atoms start at $t=0$ from the wall. We set the upper limit of the complete pulse duration to \SI{2}{\nano\second} and for that $\SI{97}{\percent}$ of all released atoms have not collided with either cell wall. Independently of the distribution we will only simulate those particles which have not collided into the cell walls during these first \SI{2}{\nano\second}. We also only take particles into account for which the time-averaged Rabi frequency is at least $\SI{10}{\percent}$ of the maximum value in the middle of the Gaussian profile. For that, we consider a small beam waist of $\mathrm{w}_{0,1}=\SI{0.5}{\micro\meter}$ for the first laser and a broad waist $\mathrm{w}_{0,2},\mathrm{w}_{0,3}=\SI{2}{\micro\meter}$ for the second and third one. This way of choosing the beam waist ensures that only particles inside of the Rydberg blockade radius are excited to the intermediate state, but each of them then is transferred from the Rydberg state to the excited state even when they fly out of the center of the beam.

Due to the Gaussian profile of the three lasers, the motion of the $N$ atoms and the distance-dependent interaction strength of the $N(N-1)/2$ particle pairs we have $3N+N(N-1)/2$ different time-dependent variables (see Eq.~\ref{eq2},\ref{eq3}). Instead of treating all those variables separately, we approximate them in time via a polynomial fit $f$ and group terms by their time order $a$, e.g. $ \sum_{n} \Omega_{1,n}(t) \vert i_n \rangle \langle g_n \vert \approx \sum_a^O (\sum_{n} f^{(a)}(\Omega_{1,n}) \vert i_n \rangle \langle g_n \vert) t^a$. We fit each variable with a polynomial up to a certain order $O$, which was $O_1=3$ for the first laser, $O_{2,3}=2$ for second and third laser, and $O_{\mathrm{int}}=10$ for the interaction energy. In this way, we can reduce the amount of time-dependent operators for the simulation remarkably.

\subsection{Effective description of the low excitation sector}\label{Effective description of the low excitation sector}
\begin{figure}[ht]
\includegraphics[width=0.79\linewidth]{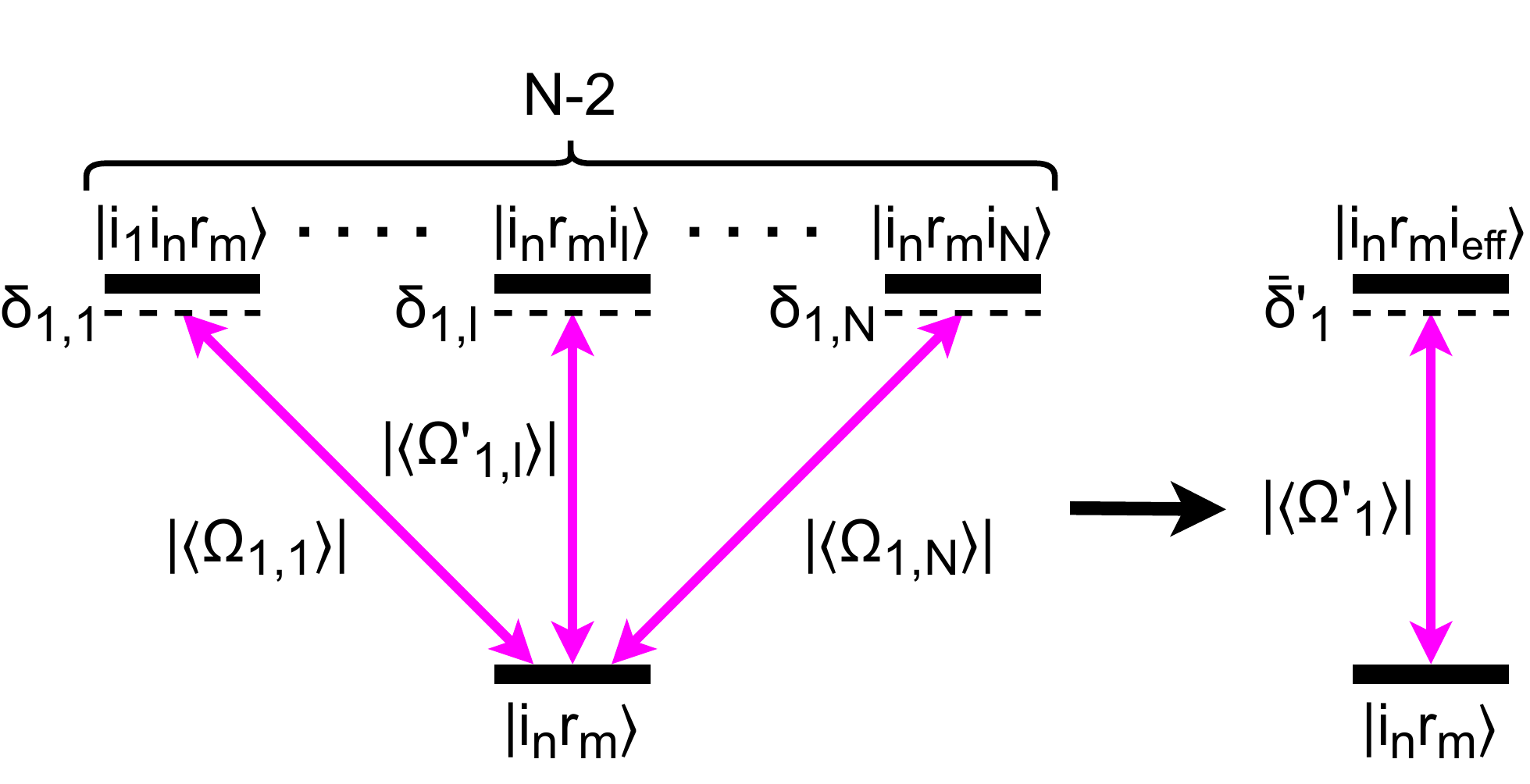}
\caption{Illustration of the effective description of the low-excitation sector. The oscillations from a state with two excitations, here $\vert i_n r_m \rangle$, to $N$-2 states with a third excitation, get combined to one oscillation to an effective state $\vert i_n r_m i_{\mathrm{eff}} \rangle$.}
\label{Effective}
\end{figure}

For the excitation process we only simulate the low excitation sector, which in our case means that we take the first two excitations fully into account and add only a third effective excitation of atoms in the intermediate state, e.g. $\vert i_n r_m i_l \rangle$, $\vert r_n r_m i_l \rangle$. All states with $\vert r_n r_m r_l \rangle$, $\vert r_n r_m e_l \rangle$ and higher will be neglected since the energy shift of three Rydberg states is much higher than the Rabi frequency of the first two lasers even for longer particle distances and so these states can never be populated for all practical purposes. Note that the notation $\vert i_n r_m i_l \rangle$ (and similar states) means that all atoms are in the ground state apart from the atoms $m$, $n$ and $l$ that are in the state indicated by the notation. We now want to describe the oscillations in Fig.~\ref{Effective}, between a state with two arbitrary excitations $n$, $m$ and the $N$-2 states with an additional atom in the intermediate state, effectively. We do this by combining the $N$-2 states into one effective state and considering the effective oscillation between this effective state and the state carrying two excitations.

For that, we will use third order perturbation theory, where we take the deviations of the Doppler shift from the mean velocity as perturbation (see Fig.~\ref{Effective} for an illustration). We define $\bar{\delta}_1=\frac{1}{N-2}\sum_{l\neq n,m} \delta_{1,l}$ and $\vert \langle \Omega_{1}\rangle \vert^2=\sum_{l\neq n,m} \vert \langle\Omega_{1,l}\rangle \vert^2$, where $ \langle \Omega_{1,l}\rangle$ are the time-averaged Rabi frequencies of each atom. Now, the eigenenergies of the Hamiltonian projected to this ($N$-1)-dimensional subspace without perturbation are
\begin{eqnarray}
E_\pm=\hbar \frac{-\bar{\delta}_1 \pm \sqrt{\bar{\delta}_1^2 + \vert \langle \Omega_{1}\rangle \vert^2}}{2}
\label{eq16}
\end{eqnarray}
and $E_{1,..,N-3}=-\hbar \bar{\delta}_1$. We then take $\bar{\delta}_1-\delta_{1,l}$ as the perturbation and calculate the corrected eigenenergies $E'_{\pm}$ to get the detuning and Rabi frequencies for the oscillations between a state with two excitations $n$, $m$ and a third effective one $\vert i_n r_m i_{\mathrm{eff}} \rangle$, see also Fig.~\ref{Effective}:
\begin{eqnarray}
&&-\bar{\delta}_1'=\frac{E_{+}'+E_{-}'}{\hbar}\\
&& \vert \langle \Omega_{1}' \rangle \vert=\frac{1}{\hbar}\sqrt{(E_{+}'-E_{-}')^2-(E_{+}'+E_{-}')^2}.
\label{eq17}
\end{eqnarray}
The $N$-3 states corresponding to the eigenenergies $E_{1,..,N-3}'$ are dark states and not further considered.

One has to mention, that we only take the time-averaged Rabi frequency for these effective oscillations for the reason of numerical simplicity. For all other transitions (involving at maximum two excitations) we still use the time-dependent Rabi frequencies.
Apart from these fast back and forth oscillations, there can be slow indirect oscillations from one state already containing an atom in the intermediate state to another. These oscillations are given by the effective Rabi frequency from the adiabatic elimination \cite{Brion2007}
\begin{eqnarray}
\hat{\tilde{H}}_{\mathrm{eff}}(t)&&=\frac{\hbar}{4} \sum_{\substack{n\\m<l}} \frac{\Omega_{1,l}(t)\Omega^*_{1,m}(t)}{\delta_{1,n}+\delta_{2,n}+\delta_{1,m}+\delta_{1,l}} 
\vert r_n,i_l \rangle \langle r_n,i_m \vert \nonumber \\
&&+h.c.
\label{eq18}
\end{eqnarray}

With this effective description of the transitions one can still capture all the relevant information of the third excitation sector with a quadratic growth of the truncated Hilbert space. The solution of the time-dependent Schrödinger equation of the excitation process is then solved with Qutip \cite{Qutip}.

\section{Results}\label{Results}
We will now present the results from the simulation and optimization of the excitation process as well as the emission properties from the decay process.
Before we optimize the laser pulse sequence we first have to determine the optimal target state $W(t_W)$  by investigating the emission pattern obtained for different $t_W$ in Sec.~\ref{Optimal Target state}. After that we optimize the pulse sequence and Rabi frequencies for maximal excitations and minimal phase differences of the singly-excited states in Sec.~\ref{Optimization of the pulse sequence}. Finally, in Sec.~\ref{Atom decay and photon emission}, we will use the final states from the simulation of the excitation process as initial states for the decay process and investigate the behavior of the single- and two-photon emission.

\subsection{Optimal target state}\label{Optimal Target state}
\begin{figure}[ht]
\begin{subfigure}{0.49\linewidth}
	\includegraphics[width=\linewidth]{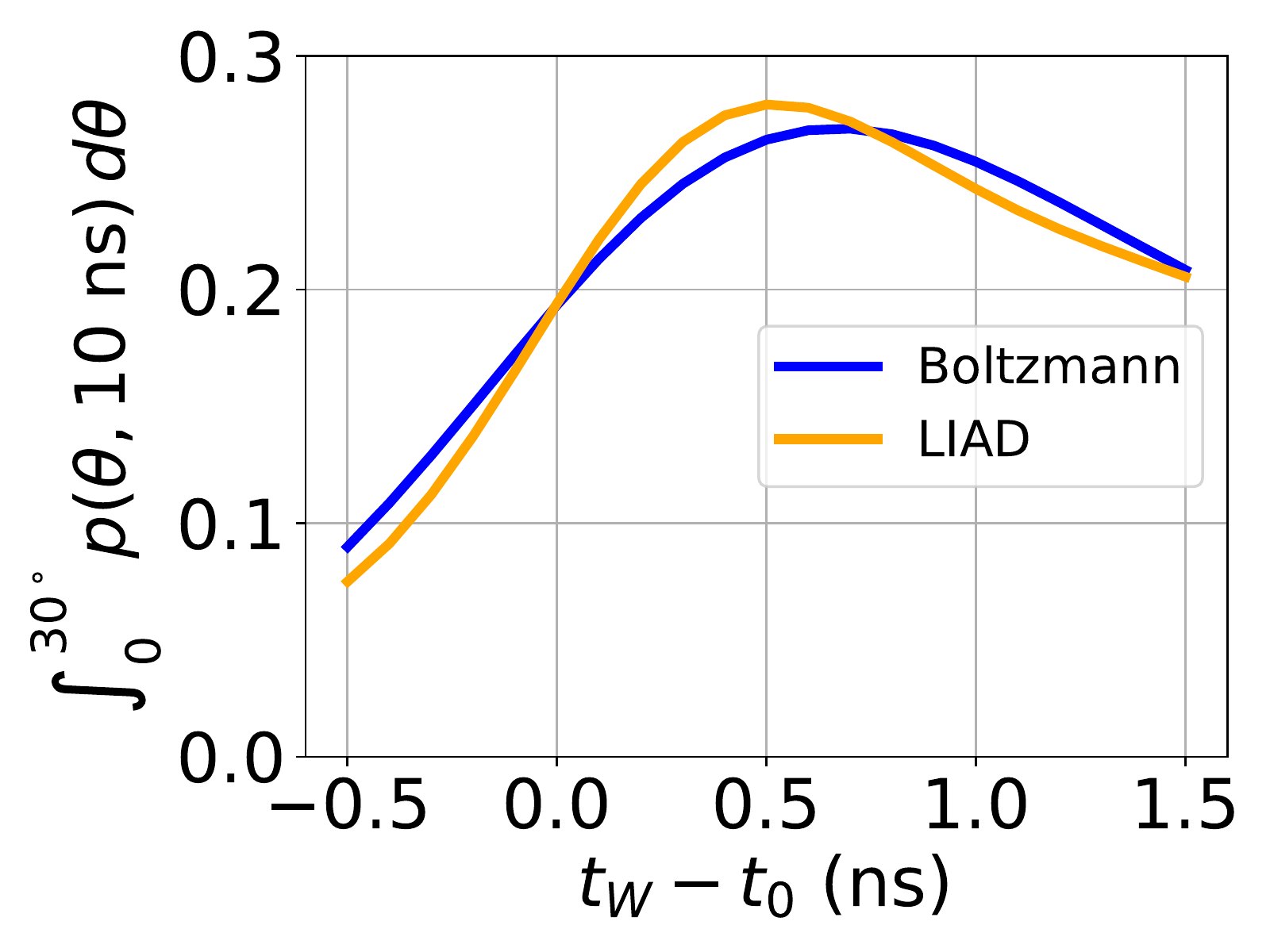}
	\caption{} \label{Wstate_P}
\end{subfigure}
\begin{subfigure}{0.49\linewidth}
	\includegraphics[width=\linewidth]{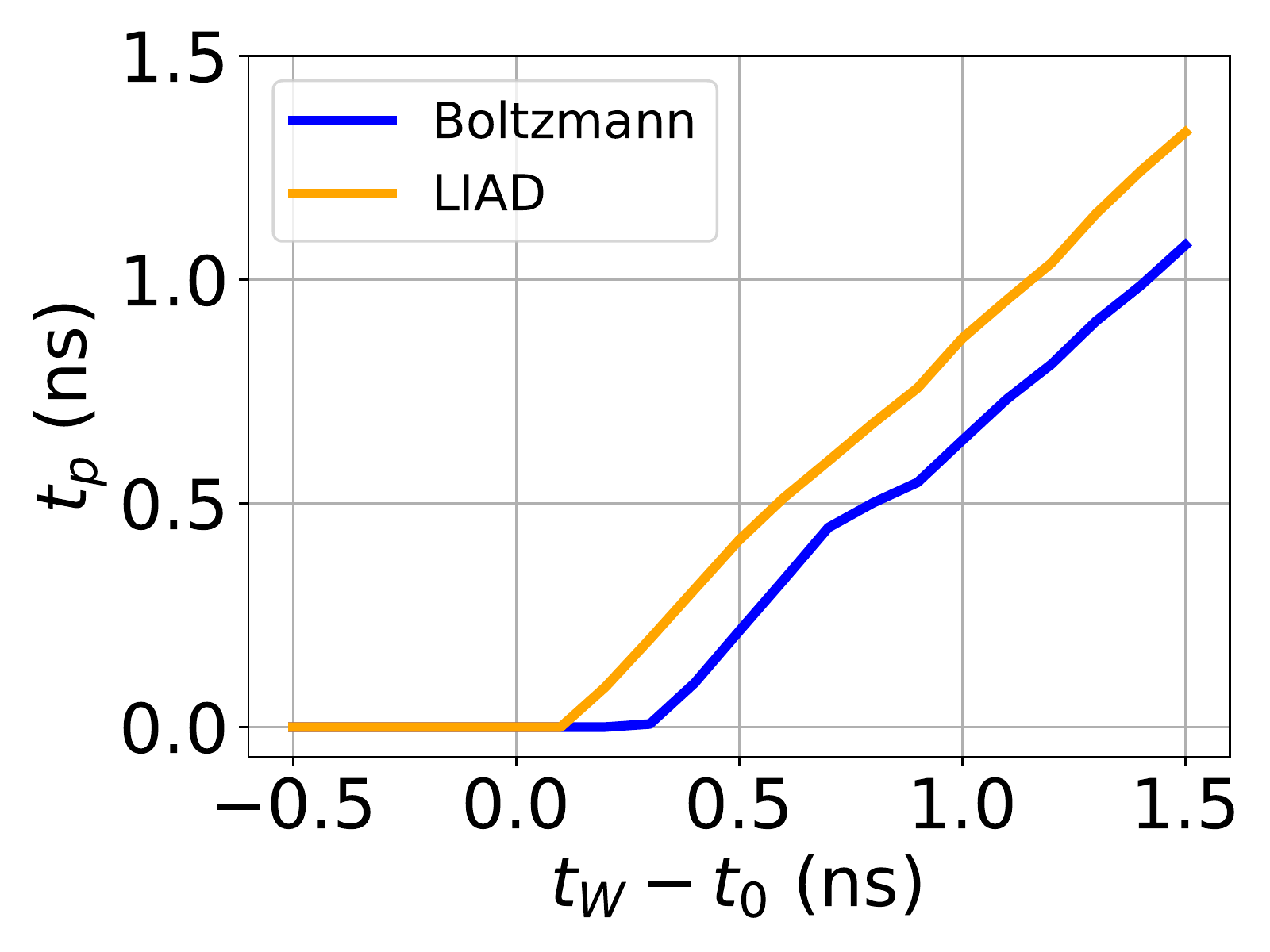}
	\caption{} \label{Wstate_ER}
\end{subfigure}
\caption{Emission properties of the target state. Total photon emission after \SI{10}{\nano\second} into a forward cone up to a \SI{30}{\degree} angle to the laser direction (a) and photon emission peak position $t_p$ (b) for different W-state phases with $N=500$ atoms averaged over 10 samples.}
\label{Wstate}
\end{figure}

In this section we will explain what would happen, if we fully reached the target state and then use this result to determine which target state has the best emission properties. We thus assume that we prepared our system with a laser pulse sequence of duration $t_0=\SI{1.5}{\nano\second}$ which drives the population into the state $\sum_{n=1}\alpha_n(t_0)|e_n\rangle$ with coefficients $\alpha_n(t_0)=\exp{ \bm(i\mathbf{k}_0 \cdot \mathbf{R}_n(t_W)\bm)}/\sqrt{N}$, where $t_W$ is a point in time at which the phase matching is maximized and the highest single-photon emission rate is roughly expected. We focus here only on the phases of the coefficients, which can vary due the motion of the atoms and the state-preparation pulses. We will discuss this in more detail in the subsequent section \ref{Optimization of the pulse sequence}. We then calculate the collective decay and photon emission via Eqs.~(\ref{eq7}) and (\ref{eq8}) for different times $t_W$ to determine the maximal photon population after several nanoseconds of decay time, even if we expect that the bulk photon population will be emitted during the first \SI{2}{\nano\second} after the laser pulse. Furthermore, we consider that all atoms which collide into the cell walls, from that moment do not contribute to the collective decay anymore, similar to the results about the finite time of flight in Ref.~\cite{Christaller2022}. Fig.~\ref{Wstate_P} shows that the maximal photon population reaches the detector if the phase of the atoms corresponds to a position $t_W$ forward in time with respect to $t_0=\SI{1.5}{\nano\second}$ (i.e., $t_W>t_0$) and that its value varies by about a factor of 3 in the analyzed range of $t_W$.
Additionally, the peak emission rate occurs at a time $t=t_0+t_p$ that is roughly proportional to $t_W-t_0$ as one can see in Fig.~\ref{Wstate_ER}.

In conclusion, we find that we get the most favorable emission behavior for $t_W\approx \SI{2}{\nano\second}$ and thus choose this value for the parameter in our target state $W(t_W)$.

\subsection{Optimization of the pulse sequence}\label{Optimization of the pulse sequence}
To optimize the pulse sequence, we first have to fix our target state $W(t_W)$. Based on the analysis of the previous section we choose $t_W=\SI{2}{\nano\second}$. Furthermore, we have to fix the weights $w_n$. Unlike in the previous section \ref{Optimal Target state}, where we chose equal weigths $1/\sqrt{N}$, we now choose the time averaged Rabi frequencies $w_n=\vert \langle \Omega_{12,n} \rangle \langle \Omega_{3,n} \rangle \vert / \sqrt{\sum_n \vert \langle \Omega_{12,n} \rangle \langle \Omega_{3,n} \rangle \vert^2}$, where $\Omega_{12,n}=\Omega_{1,n}\Omega_{2,n}/(2\delta_{1,n})$ is the Rabi frequency for the two-photon transition after adiabatic elimination of the intermediate level $|i\rangle$~\cite{Brion2007} (note that in the simulation we simulate also the intermediate level and do not actually perform the adiabatic elimination). This choice of the weights $w_n$ reflects the influence of the gaussian laser profile and ensures that the optimization focuses on the atoms in the center of the laser beam where the intensity is close to its maximum. Note, that even if equal weights may seem reasonable, they would cause the optimized Rabi frequencies to go beyond every technical limit in an attempt to excite also the atoms in the tail of the Gaussian laser-beam profile.

We can now proceed with the optimization of the pulse sequence, where we optimize in particular the position of the three laser pulses and their amplitudes, i.e., the constant Rabi frequencies. We furthermore use the total pulse duration $t_0$ as an optimization parameter, and allow values out of the interval $\SI{1.25}{\nano\second}\leq t_0\leq \SI{1.75}{\nano\second}$. We simulate 50 different samples with 30 atoms each, where we randomly sample the positions and velocities of the atoms.
We then use the mean (in the sense of an average over the 50 samples) fidelity $F_W=\vert \langle W(t_W) \vert \psi (t_0) \rangle \vert^2$ as Figure of Merit
for a maximization with the Nelder-Mead algorithm \cite{Nelder1965} implemented in the optimization software Quantum Optimal Control Suite (QuOCS)~\cite{Reisser2023}.

\begin{figure}[ht]
\includegraphics[width=0.6\linewidth]{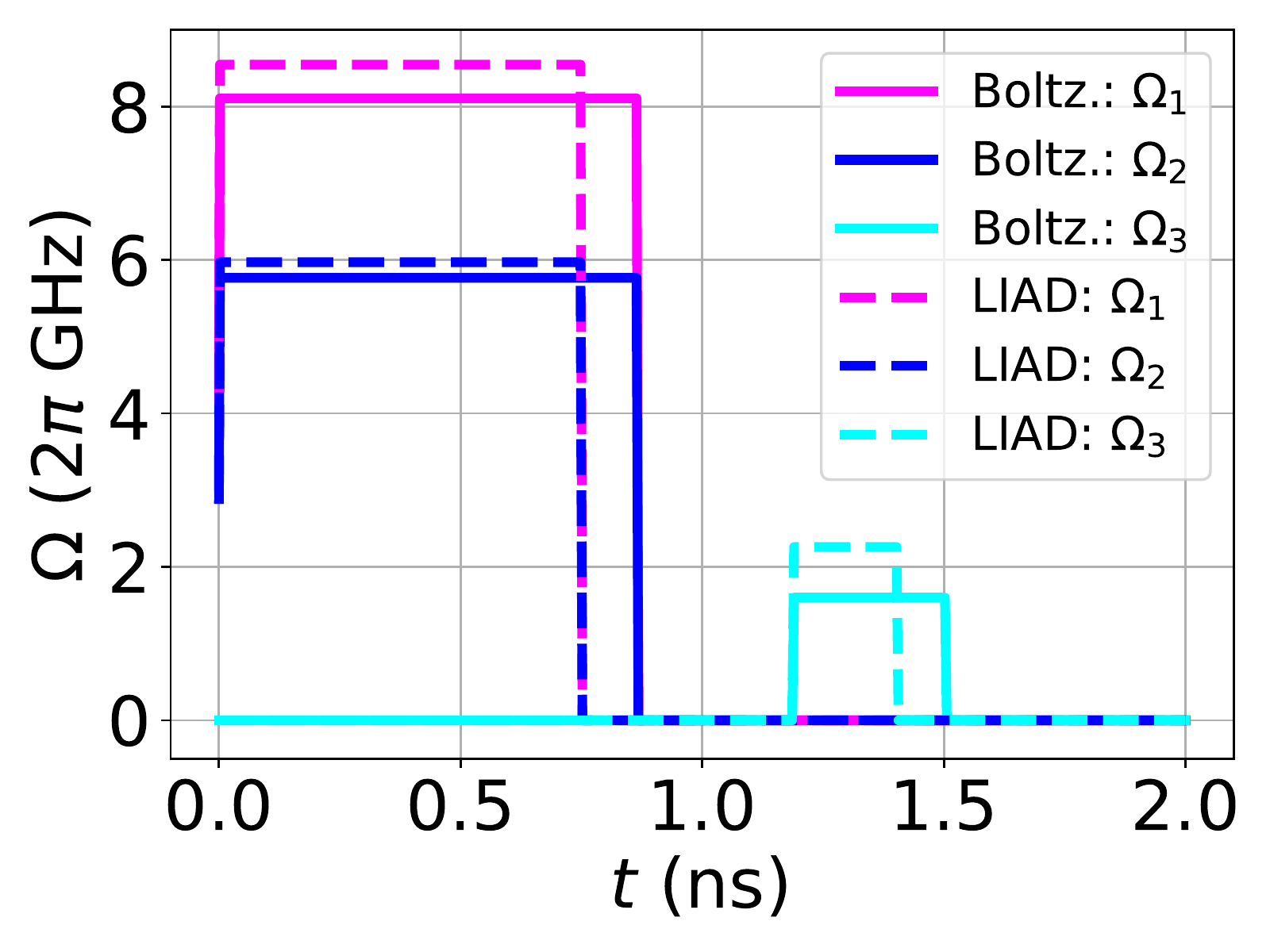}
\caption{The optimized pulses obtained from the simulation of 50 samples each with $N=30$ atoms each. The atoms were sampled once according to the Boltzmann distribution and once according to the distribution in the LIAD case and the pulses were optimized separately for all the four scenarios. The delay between the pulse of the first two lasers and the third one further increases the phases.}
\label{Pulse}
\end{figure}

The resulting pulse shapes are shown in Fig.~\ref{Pulse}. We can see a time gap between the end of the first two lasers, at $\Delta t_{12}$, and the start of the third one, at $t_{s,3}$, which resembles a spin echo pulse \cite{Hahn1950}. This can be explained by looking at the velocity-dependent phase of the excited state for a single atom $\phi_e(\mathbf{v})\approx[\mathbf{k}_0 \Delta t_{12}-\mathbf{k}_3 (2t_{s,3}+\Delta t_3-\Delta t_{12})] \cdot \frac{\mathbf{v}}{2}$ (see Appendix~\ref{Appendix B}). Since we have an anti-parallel laser geometry, $-\mathbf{k}_3$ has the same direction as $\mathbf{k}_0$, but $k_3>k_0$ and with that the second term becomes more dominant. Therefore, the corresponding phase time $t_\phi=\frac{\phi_e(\mathbf{v})}{\mathbf{k}_0 \cdot \mathbf{v}}$ can be noticeably greater than the pulse duration $t_0=t_{s,3}+\Delta t_3$ if $t_{s,3}>\Delta t_{12}$. While this calculation is for a single atom, the same consideration holds also for the many-body state where we can numerically calculate how the pulse duration and timing contribute to obtaining the phases that correspond to the parameter $t_W$ of the target state. The fidelities, the populations of the ground, singly- and doubly-excited states and the phase time difference with respect to the pulse lenght are given in Tab.~\ref{Fidelities}. The main limiting factor for the state preparation is caused by the relative phases, while the population of the singly-excited states exceeds the state preparation fidelity in all cases. Additionally, there is a residual population of the ground state resulting in a no-photon error. The most problematic error for our single-photon source is the population of the doubly-excited state of up to about $10\%$ which will lead to the emission of a second photon. In the next section, we will study the emission of the photons and see that the deleterious effect of the second photon is suppressed by its different emission properties. Nevertheless, the state preparation could probably be improved using for example more complex pulse shapes based on quantum optimal control.

\begin{table}[ht]
\caption{The fidelity $F_W$, the populations of the ground (g), singly- (e) and doubly-excited states (ee) and the phase time and its difference to the the pulse duration. The difference of the phase time with respect to the pulse length $t_\phi-t_0$ is in both cases close to the desired \SI{0.5}{\nano\second} which we took from Fig.~\ref{Wstate_P}.}
\begin{ruledtabular}
\begin{tabular}{lll}
 & Boltzmann & LIAD    \\ 
 $F_W~(\%)$ & $72.0 \pm 4.0$ & $65.5 \pm 3.4$   \\ 
 $g~(\%)$ & $3.5 \pm 3.2$ & $8.0 \pm 4.7$  \\ 
 $e~(\%)$ & $76.3 \pm 4.4$ & $68.8 \pm 3.9$  \\ 
 $ee~(\%)$ & $8.2 \pm 2.0$ & $9.0 \pm 2.2$  \\ 
 $t_\phi-t_0~(\mathrm{ns})$ & $0.468$ & $0.524$ 
\end{tabular}
\end{ruledtabular}
\label{Fidelities}
\end{table}

\subsection{Atom decay and photon emission}\label{Atom decay and photon emission}
\begin{figure}[b]
\begin{subfigure}{0.49\linewidth}
	\includegraphics[width=\linewidth]{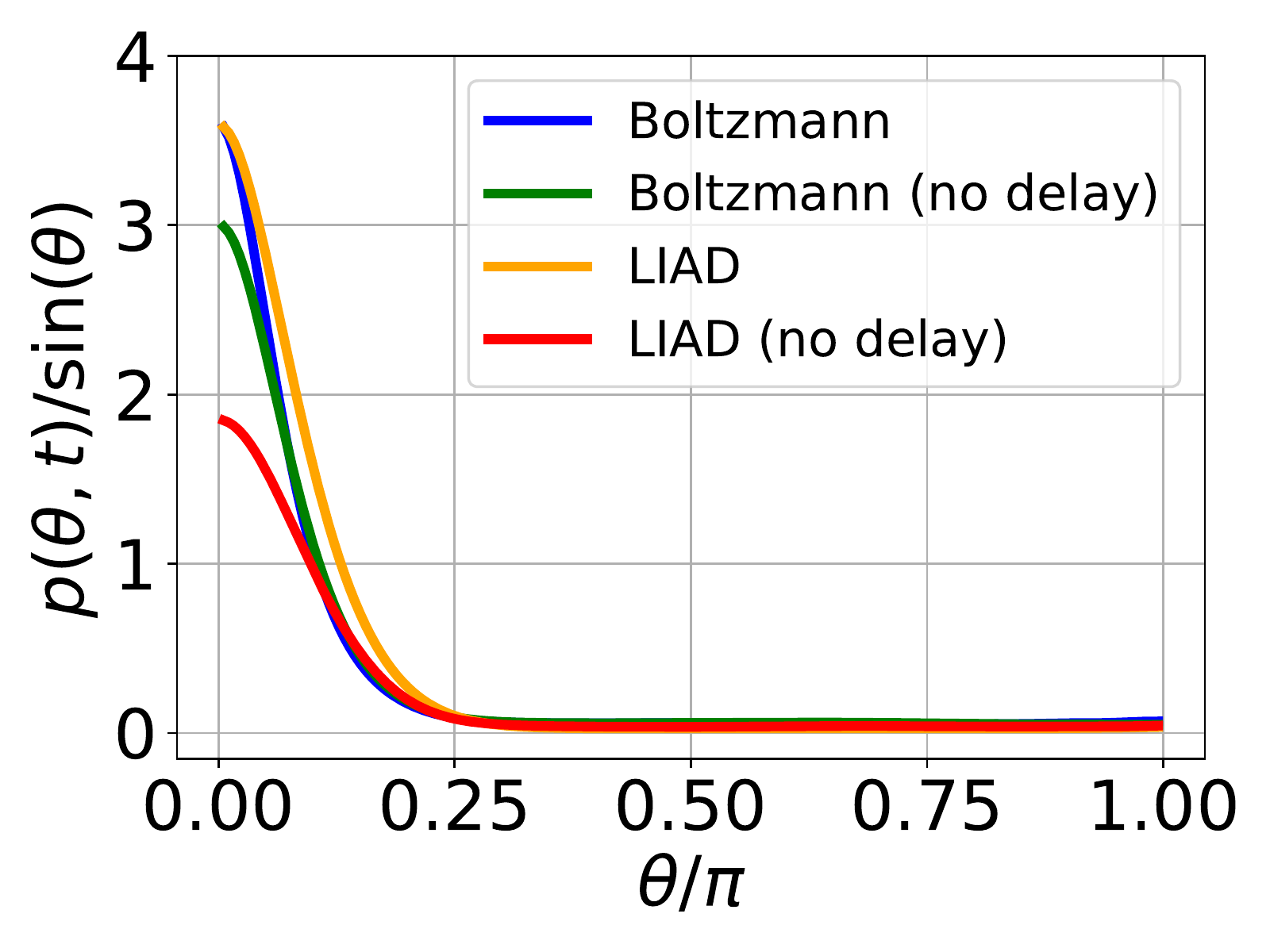}
	\caption{} \label{P_sin_theta}
\end{subfigure}
\begin{subfigure}{0.49\linewidth}
	\includegraphics[width=\linewidth]{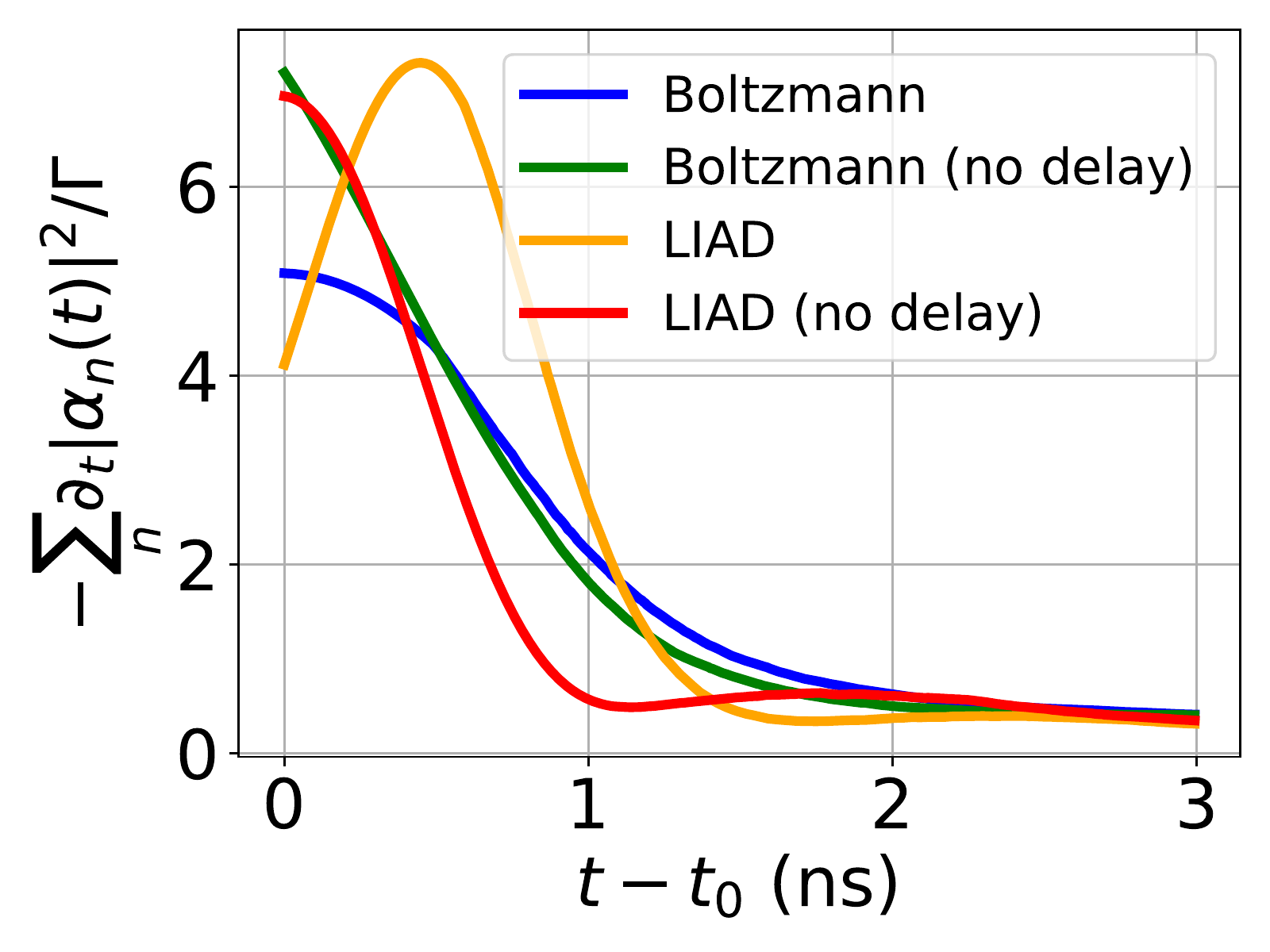}
	\caption{} \label{ER}
\end{subfigure}
\begin{subfigure}{0.49\linewidth}
	\includegraphics[width=\linewidth]{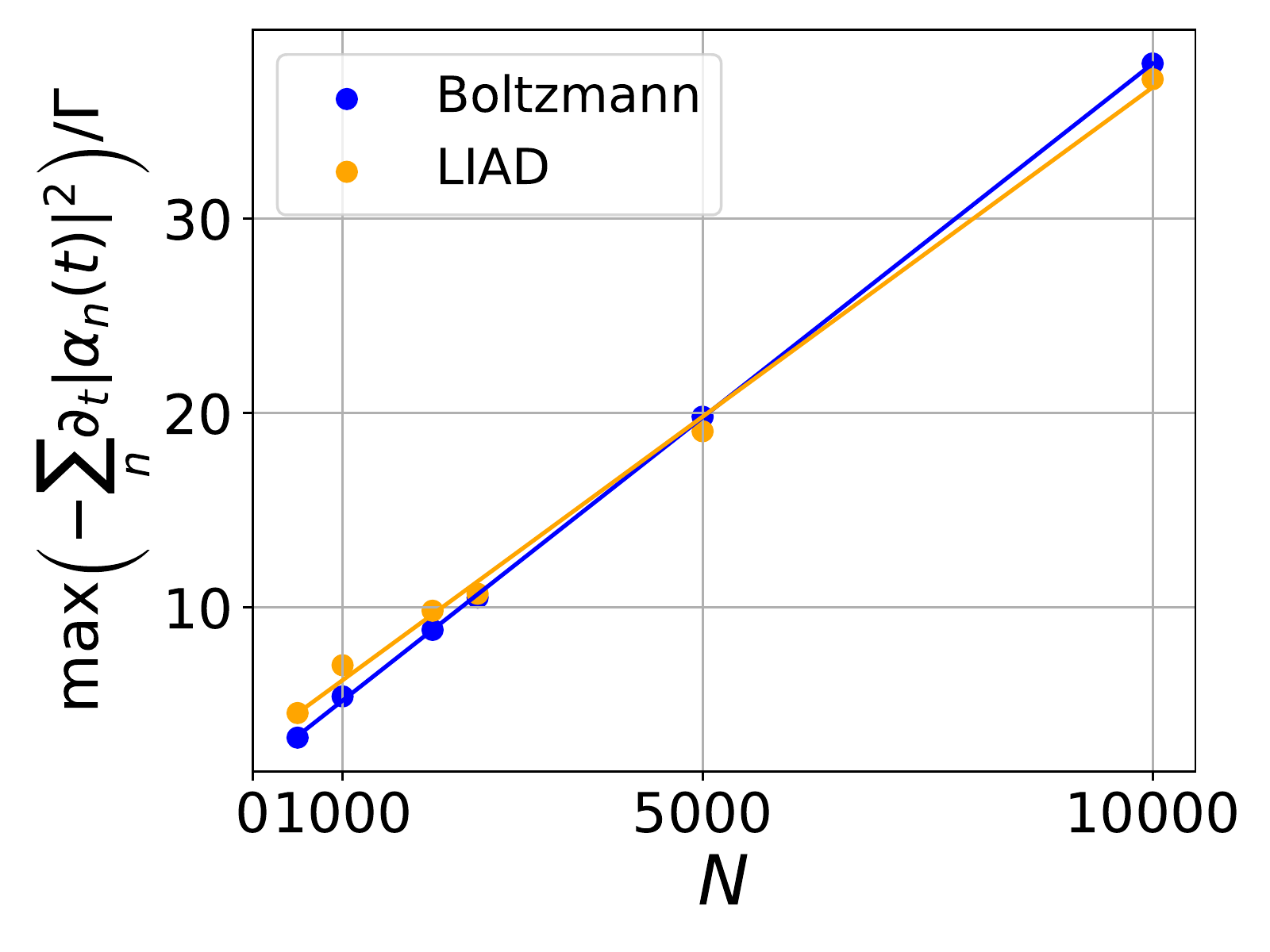}
	\caption{} \label{ER_N}
\end{subfigure}
\caption{Single-photon emission. (a) shows the photon-population density function from the decay of a single excitation after \SI{10}{\nano\second} where we excluded the geometrical factor $\sin (\theta)$. In (b) we show the associated emission rate in units of the corresponding single-atom decay rate. Both, (a) and (b), also show a comparison between the optimized pulse with the delay of the third laser and without. (c) shows a linear dependency of the maximal emission rate on the particle number $N$.}
\label{P_ER}
\end{figure}

To calculate the decay of the atomic excitation and the photon emission we first take the optimized pulses from the previous section and simulate the excitation process of $100$ samples with each $N=100$ atoms. Since in a real experiment the number of atoms is at least a factor of 10 higher and coefficients $\alpha_n(t_0)$ of the singly-excited states are nearly independent from each other (but depend on the velocity and initial position of each individual atom), one can group always 10 from the 100 samples together and only needs to divide the coefficients $\alpha_n (t_0)$ by a factor $\sqrt{10}$ to obtain a realistic estimate for the single-excitation contribution of a 1000-atoms ensemble. We can take this state as the initial state to calculate the decay process, where we thus consider 10 disjoint groups of samples with 1000 atoms each. In principal, one could do this for an arbitrary number of samples, as long as it is a divisor of the total number of samples. In Fig.~\ref{ER_N} we increased this number to the total number of samples, but for the detailed analysis of the photon direction we limit this number up to 10, since the directional photon population in Eq.~\ref{eq8} is depending quadratic on the number of particles. For the doubly-excited states we do not group the samples together, mainly, since their coefficients $\alpha_{n,m} (t_0)$ can not be factorized into two single-particle coefficients, and further because of the quadratically growing number of atom pairs $N(N-1)/2$ and the associated high computational cost; instead, we calculate the decay for all of the 100 samples separately.

\begin{figure}[ht]
\begin{subfigure}{0.49\linewidth}
	\includegraphics[width=\linewidth]{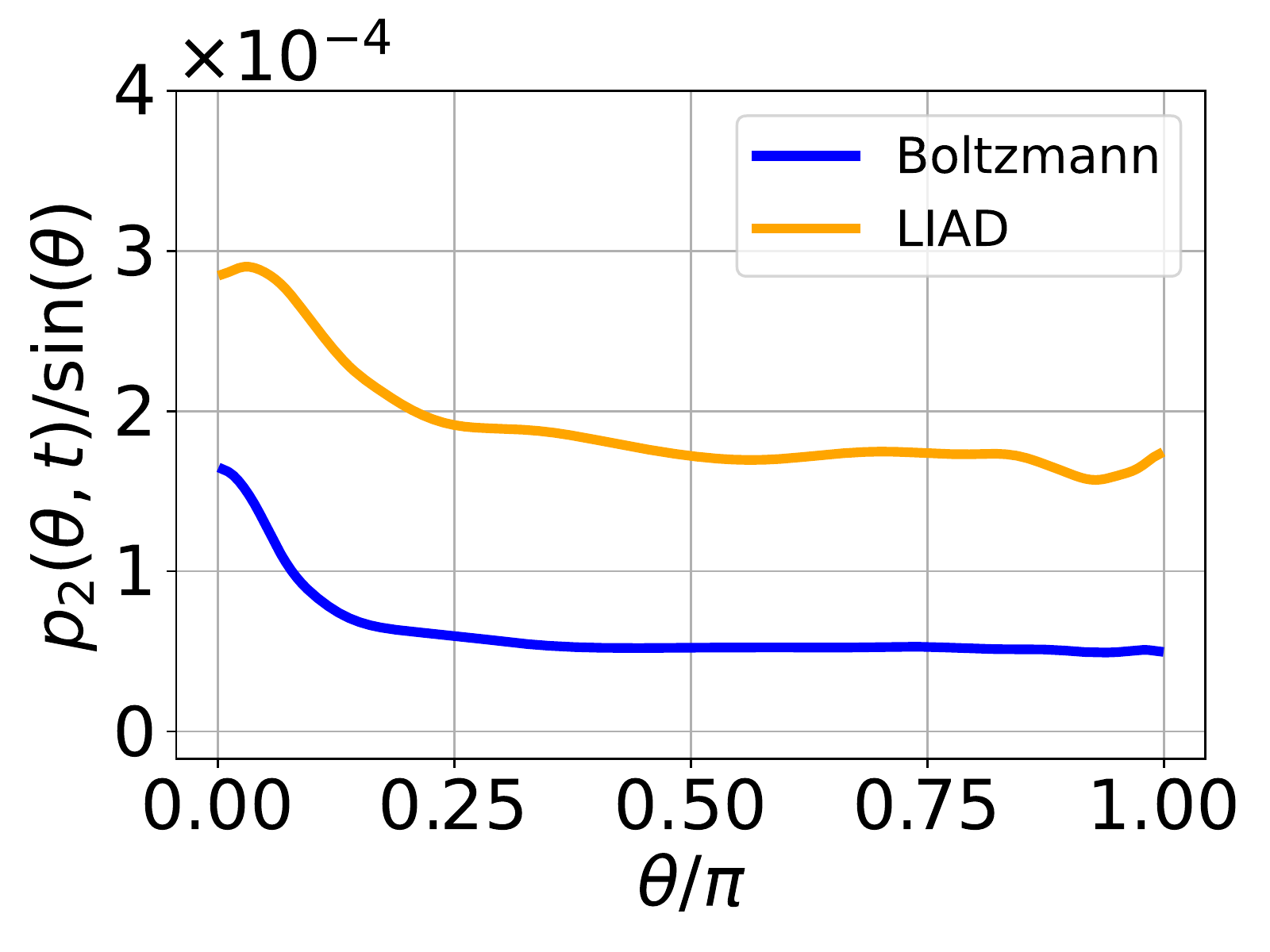}
	\caption{} \label{P2_sin_theta}
\end{subfigure}
\begin{subfigure}{0.49\linewidth}
	\includegraphics[width=\linewidth]{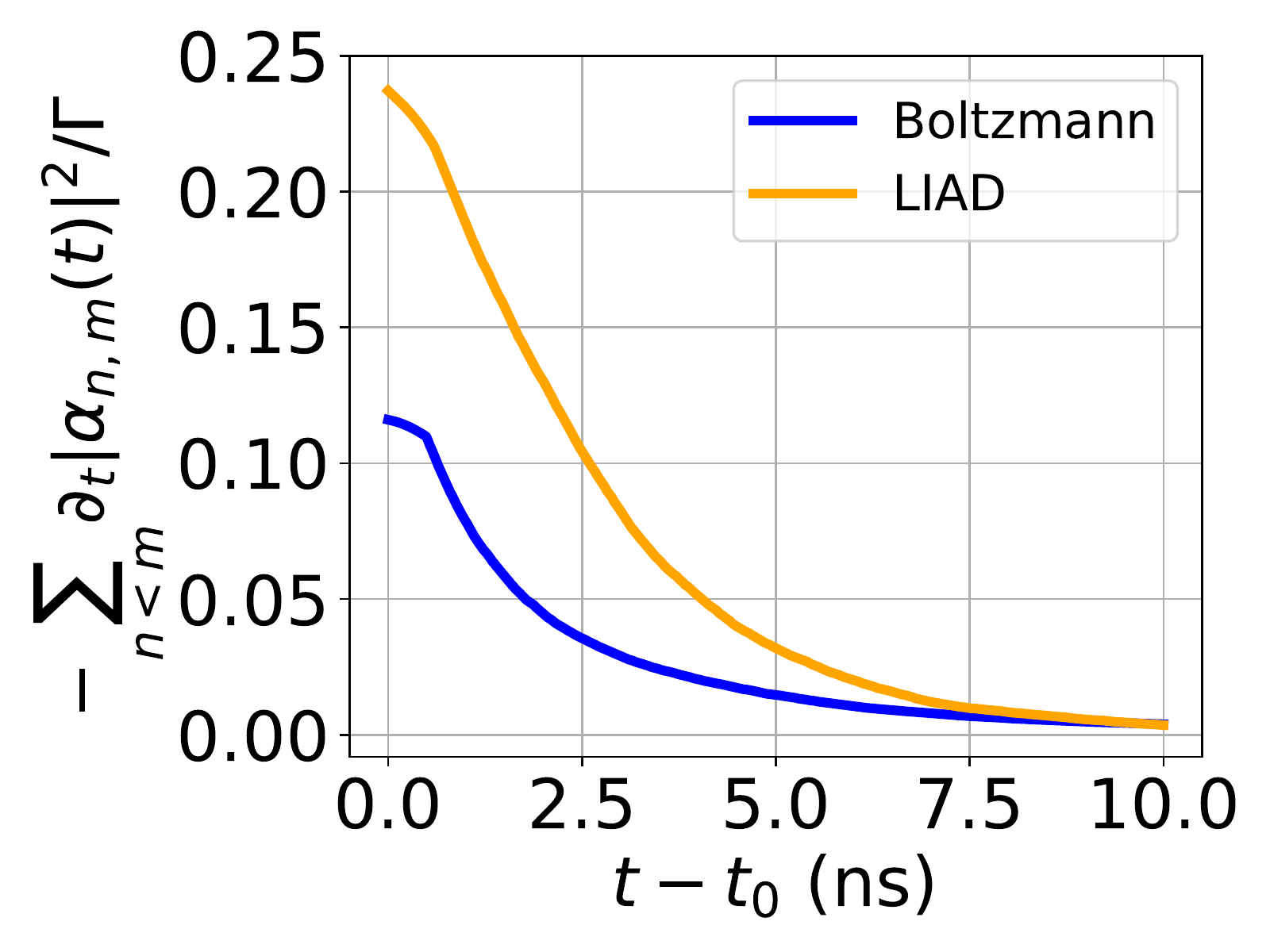}
	\caption{} \label{ER21}
\end{subfigure}
\begin{subfigure}{0.49\linewidth}
	\includegraphics[width=\linewidth]{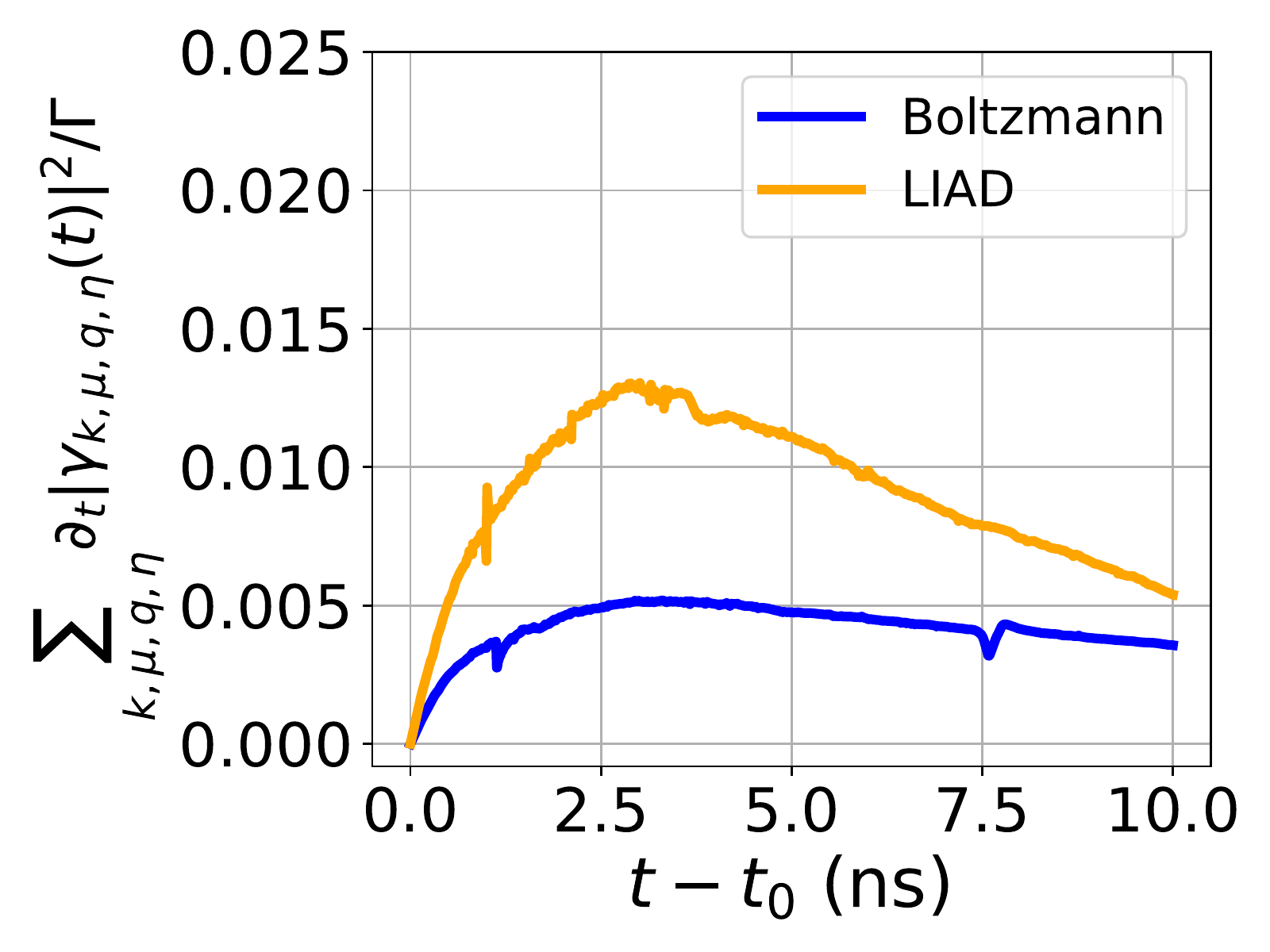}
	\caption{} \label{ER22}
\end{subfigure}
\caption{Two-photon emission. (a) shows the photon population density function for the first photon emitted from the double-excitation, while the corresponding emission rate in units of the single-atom decay rate is shown in (b). In (c) one can see the emission rate of the second photon. The second photon is emitted at a much lower rate and with a peak at larger $t$ compared to the single-photon. In the LIAD case we neglected three samples for this plot due to numerical artifacts.}
\label{P_ER2}
\end{figure}

As we can see in Fig.~\ref{P_sin_theta}, a large amount of population is emitted with a small polar angle to $\mathbf{k}_0$. The emitted population up to an angle of $\theta=30^\circ$ is in the case of Boltzmann $0.1422$ and in the case of LIAD $0.1941$. This preference in the photon direction is correlated to the occurrence of a super-radiant emission burst during the first few nanoseconds, Fig.~\ref{ER}, where the collective decay rate is several times larger than the single-atom decay rate $\Gamma$. Note that we neglected possible photon emissions induced through atom-atom and atom-wall collisions. Furthermore, we split the simulation of the excitation and the decay of the atoms, and thus, we neglected a possible photon emission during the last laser pulse. This means that a combined simulation would show also a photon emission for $t<t_0$. Nevertheless, for $t>t_0$ the emission rate should look qualitatively the same.
Additionally, to clearly see the influence of the phases of the singly-excited states, we also run the simulations in Fig.~\ref{P_sin_theta},~\ref{ER} with the same pulse but without the delay of the third laser. This choice should have nearly no influence on the population of the singly-excited states at the end of the pulse, since the only time dependent part of the Hamiltonian for the excitation process is the interaction energy. The difference of the phase time corresponding to the phase information of the singly-excited state with respect to the pulse length $t_\phi-t_0$ is now in this case $\SI{0.247}{\nano\second}$ instead of $\SI{0.468}{\nano\second}$ for Boltzmann and $\SI{0.223}{\nano\second}$ instead of $\SI{0.526}{\nano\second}$ for LIAD.
As a consequence we expect that the emission pattern is changed. Indeed, we observe in the simulation that the collective emission rate changed in both cases (see Fig.~\ref{ER}). While for Boltzmann, the peak primarily got stronger at $t_0$, in the case of LIAD it moved from around $\SI{0.5}{\nano\second}$ to $t_0$. Nevertheless, for both distributions we still get a higher photon population for small angles $\theta$ if we apply the delay, especially in the case of LIAD. The differences between the Boltzmann and LIAD distributions in the photon emission, originate from the non-isotropic motion of atoms in the LIAD case, which are moving mainly parallel or antiparallel to the direction of the lasers. This seems to be the reason for the increased sensitivity to the phase information as one can see in Fig.~\ref{P_sin_theta}, \ref{ER}. We could probably amplify this effect by preparing the cell in a way, where only one of the walls emits atoms (thus reducing the relative motion among the atoms even further).\\
In Fig.~\ref{ER_N} we plotted the maximum value of the emission rate for different sample group sizes between $5$ and $100$ samples, each with 100 atoms. One can see, that the superradiant emission burst increases linearly with N at a rate of $3.64\Gamma$ (Boltzmann) and $3.39\Gamma$ (LIAD) per $1000$ atoms. This emission enhancement with the number of atoms is a promising behavior for a possible experiment, since we expect to have $N=1000$-$10000$ atoms in a real vapor cell. The exact number then depends on the temperature of the vapor cell and in the case of LIAD also on the off-resonant laser and can in principle be controlled.\\
In Fig.~\ref{P2_sin_theta} we show the photon population density function $p_2(\theta)$, Eq.~(\ref{eq9}), for the first photon from the decay of the double-excitation. Compared to the one for a single-photon, it is very close to an isotropic radiation source independent from the choice of the states. This seems not surprising, since the interaction energy results in an additional phase, which does not depend on the position and velocity of each atom on their own. With that, most of the stored information of the absorbed photons is lost through this interaction-dependent phase. One can also see that the photon emission rate is similar to an exponential function which one would get without any collective effects, Fig.~\ref{ER21}.
The emission rate of the second photon is shown in Fig.~\ref{ER22}. It is very similar to what one would expect for two separately excited atoms both with an independent decay: one would get an emission rate proportional to $\Gamma(1-\exp (-\Gamma t))\exp (-\Gamma t)$. One reason that the emission peak obtained from the simulation occurs slightly earlier than $\ln(2)/\Gamma$, is (apart from a slight collective behavior) the fact that we set the condition that an atom can only contribute to the collective decay before it collides with the wall. If we compare the results for LIAD and Boltzmann for the two-photon emission, there seems to be no qualitatively difference, in contrast to the single-photon emission. This is likely due to the scrambled phase information, which gets destroyed in both cases by the high interaction energy.

In summary, the decay of the single-excitation can show strong collective effects on the photon direction and emission rate, while the decay of the double-excitation doesn't show these collective properties.

\section{Conclusion}\label{Conclusion}
We have studied numerically the full dynamics of an on-demand single-photon source based on a room-temperature ensemble of Rubidum atoms in a micro cell. We have divided the process into the excitation and decay processes which allowed us to apply appropriate approximations for the two regimes. We have analyzed the influence of the laser pulses on the phase information of the absorbed photons and as a consequence on the directionality of the emitted photon and we have optimized the pulses accordingly. At the same time we have found that the choice of position and velocity distribution of the atoms had only a marginal effect on the photon emission.

Furthermore, we have studied the impact of the phases of the singly-excited states and have shown that it is possible to encode a phase time $t_\phi$ greater than the pulse duration $t_0$ into the collective excited state (see Tab.~\ref{Fidelities}). This extends the time interval of directed emission by slowing down the motional dephasing of the state similar to a spin echo.
We have extended existing treatments of the decay process for a Rydberg-based single-photon source to include more rigorously the effect of motion of the atoms and described also the emission of a second photon from a residual double-excitation.
For all investigated configurations we have found that the double-excitation carries only negligible phase information and leads to an almost isotropic emission pattern. Together with the already low population of the double-excitation sector after the laser pulses, this undirected emission of the second photon further improves the quality of the single-photon source.

We believe that this theoretical work can help us in a planned experiment, especially towards increasing the single-photon efficiency and distinguishing single-photon emission from multi-photon emission. One could reach more realistic results by combining the simulation of the excitation process with the atomic decay. Furthermore, it might be advantageous to use optimal control and time-modulated pulse shapes to reduce the effects of the Doppler shift, the Gaussian laser profile, and time-dependent interaction energy.

\begin{acknowledgments}
We thank Matteo Rizzi, Felix Motzoi, and Niklas Tausendpfund for fruitful discussions.
This work was funded by the German Federal Ministry of Education and Research through the program ``Quantum technologies - from basic research to market'' under the project FermiQP (13N15891) and ``AIDAS - AI, Data Analytics and Scalable Simulation'', which is a Joint Virtual Laboratory gathering the Forschungszentrum Jülich (FZJ) and the French Alternative Energies and Atomic Energy Commission (CEA).
This work is also supported by the Deutsche Forschungsgemeinschaft (DFG) via Grant No. LO 1657/7-1 under DFG SPP 1929 GiRyd.\\
\end{acknowledgments}

\appendix
\section{Photon population density function}\label{Appendix A}
This Appendix contains a more detailed derivation of Eq.~(\ref{eq8}) from Sec.~\ref{Single-exitation and single-photon emission}. For more details see also Ref.~\cite{Reuter2022}.
For the spatial profile of the single-photon emission, we are interested in the probability of the wave vector $\mathbf{k}$ having an angle $\theta$ with respect to 
$\mathbf{k}_0$. We calculate the probability density function $p(\theta,t)$ by starting with the summed absolute value squared of Eq.~(\ref{eq6})
\begin{widetext}
\begin{equation}
\sum_{\mathbf{k},\mu} \vert \beta_{\mathbf{k},\mu}(t) \vert^2 = \frac{V}{(2\pi)^3} \sum_{n,m} \int\limits_{t_0}^t \int\limits_{t_0}^t \alpha_n(t') \alpha_m^*(t'') \left( \int \vert \mathrm{g}_{\mathbf{k}} \vert^2 e^{-i\mathbf{k} \cdot (\mathbf{R}_n(t')-\mathbf{R}_m(t''))}e^{i(\omega_k-\omega_e)(t'-t'')} d^3k \right) dt' dt''
\label{eq19}
\end{equation}
\end{widetext}
We can reduce the double time integral to a single one by applying the Wigner-Weisskopf approximation before we integrate over the directions. Notice, that even if the approximation disregards any contribution with $t' \neq t''$, the structural information stays the same, since the approximation is only affecting the integral over the wave number, but not over the direction of the wave vector, which can be parametrized by the angles $\theta$ and $\phi$. We then get the following form Eq.~(\ref{eq20}) with $\mathbf{d}_{n,m}^{\; \parallel}(t')$ and $\mathbf{k}_e^{\, \parallel}(\theta)$ being the parallel parts of $\mathbf{d}_{n,m}(t')$ and $\mathbf{k}_e(\theta,\phi)$ with respect to $\mathbf{k}_0$. We also set the angle $\phi$ for every particle pair as the angle between $\mathbf{k}_e^\perp(\theta,\phi)$ and $\mathbf{d}_{n,m}^\perp(t')$
\begin{widetext}
\begin{eqnarray}
\sum_{\mathbf{k},\mu} \vert \beta_{\mathbf{k},\mu}(t) \vert^2 &&\approx \frac{\Gamma}{4 \pi} \sum_{n,m} \int\limits_{t_0}^t 
\alpha_n(t') \alpha_m^*(t') \left( \int\limits_0^\pi \int\limits_0^{2\pi} e^{-i\mathbf{k}_e(\theta, \phi) \cdot \mathbf{d}_{n,m}(t')} 
\, d\phi \, \sin(\theta) \, d\theta \right) dt'\\
&&=\frac{\Gamma}{4 \pi} \sum_{n,m}  \int\limits_{t_0}^t \alpha_n(t') \alpha_m^*(t') \left( \int\limits_0^\pi  
e^{-i\mathbf{k}_e^{\, \parallel}(\theta) \cdot \mathbf{d}_{n,m}^{\; \parallel}(t')} \int\limits_0^{2\pi} e^{-i\mathbf{k}_e^\perp(\theta,\phi) \cdot 
\mathbf{d}_{n,m}^\perp(t')} \, d\phi \, \sin(\theta) \, d\theta \right) dt'.
\label{eq20}
\end{eqnarray}
\end{widetext}
Finally in Eq.~(\ref{eq21}), we obtain an integral over $\theta$ of the density function, where we insert $\mathbf{k}_e^\perp(\theta,\phi) \cdot \mathbf{d}_{n,m}^\perp(t')=k_e \sin (\theta) d_{n,m}^\perp \cos (\phi)$ into Eq.~(\ref{eq20}) and use the first kind Bessel function $J_0(\varphi)=\frac{1}{2\pi}\int_0^{2\pi}e^{-i\varphi \cos (\phi)} d\phi$
\begin{widetext}
\begin{eqnarray}
\sum_{\mathbf{k},\mu} \vert \beta_{\mathbf{k},\mu}(t) \vert^2 && \approx \int \limits_0^\theta \frac{\Gamma}{2} \sin(\theta) \int\limits_{t_0}^t \sum_{n,m} \alpha_n(t') \alpha_m^*(t')  e^{-ik_e d_{n,m}^{\; \parallel}(t') \cos(\theta)}
J_0\bm(k_e d_{n,m}^\perp(t') \sin(\theta)\bm) \, dt' \, d\theta\\
&&=\int \limits_0^\theta p(\theta,t) \, d\theta,
\label{eq21}
\end{eqnarray}
\end{widetext}
where the kernel of the equations yields the expression Eq.~\eqref{eq8} for $p(\theta,t)$.
The same calculation holds for the first photon in the case of double excitations, but instead of calculating $\sum_{n,\mathbf{k},\mu} \vert \beta_{n,\mathbf{k},\mu}(t) \vert^2$, we need to calculate $\sum_{n,\mathbf{k},\mu} \int \limits_{t_0}^t (\partial_{t'} \vert \beta_{n,\mathbf{k},\mu}^\alpha(t') \vert^2) \cdot \vert \beta_{n,\mathbf{k}}^\gamma(t') \vert^2 \, dt'$ to get Eq.~(\ref{eq9}).

\section{Velocity-dependent phase}\label{Appendix B}
The velocity-dependent phase of the singly-excited states for several atoms, which we mentioned in section~\ref{Optimization of the pulse sequence}, can be derived from the calculation of a single one. For that, we will now shortly recap the time evolution of a single atom with ground state $\vert g \rangle$ and excited state $\vert e \rangle$. The transition is driven by a laser with Rabi frequency $\Omega$ and detuning $\delta$. Furthermore, we make the exception that we start in the ground state at time $t_s$. The Hamiltonian looks like
\begin{eqnarray}
\hat{\tilde{H}}=-\hbar \delta \vert e \rangle \langle e \vert + \frac{\hbar}{2} (\Omega \vert e \rangle \langle g \vert + \Omega^*\vert g \rangle \langle e \vert).
\label{eq22}
\end{eqnarray}
The time evolution is then given by a unitary operator $\hat{\tilde{U}}(\Delta t)=\exp (-\frac{i\hat{\tilde{H}}\Delta t}{\hbar})$ which takes a simple form by writing it in Pauli matrices. After we transform the system back into the lab frame by applying $\exp (-i\delta \Delta t \vert e \rangle \langle e \vert)$,  the final state is
\begin{eqnarray}
\vert \psi (t) \rangle &&= \left[\cos \left(\frac{\Omega_\delta}{2} \Delta t \right) -i \frac{\delta}{\Omega_\delta}\sin \left(\frac{\Omega_\delta}{2} \Delta t\right)  \right]
e^{i\frac{\delta}{2} \Delta t} \vert g \rangle \notag \\
&&-i \frac{\Omega}{\Omega_\delta}\sin \left(\frac{\Omega_\delta}{2} \Delta t\right) e^{-i\frac{\delta}{2} \Delta t} \vert e \rangle
\label{eq23}
\end{eqnarray}
with the effective Rabi frequency $\Omega_\delta=\sqrt{\vert \Omega \vert^2 + \delta^2}$. If we now set the detuning equal to $\delta=-\mathbf{k} \cdot \mathbf{v}$ and the phase of the Rabi frequency as $\Omega=\vert \Omega \vert \exp \bm(i \mathbf{k} \cdot (\mathbf{R}(0)+\mathbf{v}t_s)\bm)$ we can see that the velocity dependent phase is given by
\begin{eqnarray}
\phi_e(\mathbf{v})=\mathbf{k} \cdot \mathbf{v}(t_s+\frac{\Delta t}{2})=\frac{\mathbf{k} \cdot \mathbf{v}}{2}(2t_s+\Delta t).
\label{eq24}
\end{eqnarray}
This relation is also valid for multiple atoms with more than two levels.

\bibliography{Paper}

\end{document}